\documentclass[%
 preprint,
 amsmath,amssymb,
 aps,
]{revtex4-2}

\usepackage{graphicx}

\usepackage{dcolumn}
\usepackage{amsmath,bm}
\usepackage{epstopdf,epsfig}
\usepackage{newtxtext}
\usepackage{newtxmath}
\usepackage{natbib}
\graphicspath{{Figures/}}
\usepackage{hyperref}
\hypersetup{
	colorlinks = true,
	urlcolor   = blue,
	citecolor  = black,
}

\newcommand{\RomanNumeralCaps}[1]
 \linenumbers

\usepackage{subcaption}
\usepackage{placeins}

\newcommand{\hq}{\bar{\bf q}}
\newcommand{\bn}{\bar{\eta}}
\newcommand{\al}{\alpha}
\newcommand{\mE}{\mathsfbi{E}}
\newcommand{\Gm}{G_{\text{max}}}
\newcommand{\stau}{\underline{\underline{\tau}}}
\newcommand{\Rb}{\mathcal{R}_\text{int}}
\newcommand{\p}{\partial}

\DeclareOldFontCommand{\rm}{\normalfont\rmfamily}{\mathrm}
\DeclareOldFontCommand{\sf}{\normalfont\sffamily}{\mathsf}
\DeclareOldFontCommand{\tt}{\normalfont\ttfamily}{\mathtt}
\DeclareOldFontCommand{\bf}{\normalfont\bfseries}{\mathbf}
\DeclareOldFontCommand{\it}{\normalfont\itshape}{\mathit}
\DeclareOldFontCommand{\sl}{\normalfont\slshape}{\@nomath\sl}
\DeclareOldFontCommand{\sc}{\normalfont\scshape}{\@nomath\sc}
\DeclareRobustCommand*\cal{\@fontswitch\relax\mathcal}
\DeclareRobustCommand*\mit{\@fontswitch\relax\mathnormal}
\DeclareMathAlphabet{\mathsfbi}{OT1}{\sfdefault}{bx}{sl}

\begin{document}

\preprint{APS/123-QED}

\title{On the liquid film instability of an internally coated horizontal tube}

\author{Shahab Eghbali}
 \email{shahab.eghbali@epfl.ch}
\author{Yves-Marie Ducimeti\`{e}re}%
\author{Edouard Boujo}
\author{Fran\c cois Gallaire}
\affiliation{%
 Laboratory of Fluid Mechanics and Instabilities, \'Ecole Polytechnique F\'ed\'erale de Lausanne, Lausanne CH-1015, Switzerland
}%

\begin{abstract}
We study numerically and theoretically the gravity-driven flow of a viscous liquid film coating the inner side of a horizontal cylindrical tube and surrounding a shear-free dynamically inert gaseous core. The liquid-gas interface is prone to the Rayleigh-Plateau and Rayleigh-Taylor instabilities. Here, we focus on the limit of low and intermediate Bond numbers, $Bo$, where the capillary and gravitational forces are comparable and the Rayleigh-Taylor instability is known to be suppressed.
We first study the evolution of the axially invariant draining flow, initiating from a uniform film thickness until reaching a quasi-static regime as the bubble approaches the upper tube wall. We then investigate the flow's linear stability within two frameworks: frozen time-frame (quasi-steady) stability analysis and transient growth analysis.
We explore the effect of the surface tension ($Bo$) and inertia (measured by the Ohnesorge number, $Oh$) on the flow and its stability.
The linear stability analysis suggests that the interface deformation at large $Bo$ results in the suppression of the Rayleigh-Plateau instability in the asymptotic long-time limit.
Furthermore, the transient growth analysis suggests that the initial flow evolution does not lead to any considerable additional amplification of initial interface perturbations, {\it a posteriori} rationalising the quasi-steady assumption.
The present study yields a satisfactory prediction of the stabilisation threshold found experimentally by~\citet{Duclaux2006}.
\end{abstract}

\maketitle

\section{Introduction} \label{sec:intro}
The gravity-driven flow of a viscous liquid film coating inside a solid cylinder has received attention by virtue of its rich dynamics and numerous applications.
One major industrial application of such a flow is in two-phase heat exchangers, like in evaporators~\citep{O'Neill2020} and vertical tube condensers~\citep{Revankar2005}, where the dynamics of the gas plugs and capillary blockage affect the heat transfer efficiency and pressure loss~\citep{Dobson1998,Teng1999}. 
It is also relevant in the airways of the human lung where the coupling between the moving wall of the airways and the liquid film coating the wall can cause airway closure~\citep{Heil2008,Bian2010,Levy2014}.

\indent The liquid-gas interface is prone to different instabilities due to surface tension, shear, gravity, and inertia, which eventually lead to the emergence of various patterns~\citep{Eggers2008,Gallaire2017}.
For instance, the long column may break apart into distinct plugs separated axially by collars, thus minimising the surface energy. This phenomenon is classified in a large family of hydrodynamic instabilities known as the Rayleigh-Plateau instability~\citep{Plateau1873, Rayleigh1878}. 
Flow characteristics of the viscous film coating the inner side of a tube depend also on its orientation with respect to gravity.
In the case of a vertical tube, where gravity drives an axial flow, the liquid interface is mainly destabilised by the Rayleigh-Plateau instability~\citep{Goldsmith1963}.
Different stages of such instability have been investigated through a large number of numerical and experimental studies.~\citet{Goren1962} showed analytically that in the absence of inertia, the length of plugs is set by the interface radius.~\citet{Frenkel1987} demonstrated that despite the growth of small disturbances, nonlinear saturation of instabilities can avoid the rupture of the liquid film in a certain range of axial flow parameters.
Using the long-wavelength approximation,~\citet{Camassa2014,Camassa2016} showed the existence of non-trivial axially traveling waves along the interface when the film thickness exceeds a critical value. Moreover, these studies describe how plugs form via Hopf bifurcation as the waves grow, and explore experimentally the absolute/convective properties of the traveling waves.
With this general picture, further studies address the effects of flow parameters on the dynamics of the flow instability, e.g. air driven flow~\citep{Camassa2017} and multiple liquid layers~\citep{Ogrosky2021v}, wall porosity~\citep{Liu2017}, Marangoni effect~\citep{Ding2018}, and the presence of surfactant~\citep{Ogrosky2021}.

Alternatively, in the case of liquid film coating the inner side of long horizontal or inclined tubes, where gravity is not orthogonal to the cross-section of the tube, a second instability, the Rayleigh-Taylor instability~\citep{Rayleigh1882,Taylor1950}, may arise.
As a result, the heavy liquid film accelerates into the light gaseous core in the direction of the gravitational field, thus forming suspended droplets, moving lenses, or rivulets~\citep{Trinh2014,Balestra2016,Balestra2018}, extending the case of a flat overhanging thin liquid film~\citep{Fermigier1992}. 
Therefore, in the case of a horizontal tube, both Rayleigh-Plateau and Rayleigh-Taylor instabilities may potentially coexist depending on the film thickness~\citep{Benilovetal2005,Benilov2006}.
~\citet{Trinh2014} investigated both experimentally and theoretically, in the absence of any axial flow, the stability of a thin viscous film that coats the underside of a tube.
Focusing on the situations where gravitational forces dominate surface tension forces, they did not evidence any manifestation of the Rayleigh-Plateau instability, highlighting that finite wall curvature and sufficiently high surface tension can suppress the Rayleigh-Taylor instability as the liquid drains.
Using a similar flow configuration, and also employing the lubrication approximation,~\citet{Balestra2016} showed that when gravity overcomes capillarity, the linear transient growth of the disturbances can lead to the formation of tiny spanwise homogeneous traveling waves at the top of the interface, which may pinch or decay as they travel downward.
\citet{Balestra2018} however showed that spanwise periodic (but streamwise homogeneous) structures called rivulets display significantly larger transient gains, rationalising their experimental observations.

\indent By increasing the film thickness and/or lowering the tube curvature, liquid drainage is enhanced and the Rayleigh-Taylor instability is dampened, thus allowing the Rayleigh-Plateau instability to occur.
Through an experimental investigation of a wider range of film thicknesses,~\citet{Duclaux2006} evidenced that when capillary forces dominate over gravity forces, the Rayleigh-Plateau instability sets in, but when they stop to dominate, the Rayleigh-Plateau instability can be suppressed by gravity.
In this case, liquid drainage is faster than capillarity-induced interface undulations and plug formation. 
Hence, gravity and surface tension play dual roles in interface destabilisation. While gravity promotes the Rayleigh-Taylor instability at the top of the interface, it opposes the formation of hanging droplets by enhancing drainage progressively while draining down the curved substrate.
In contradistinction, surface tension promotes the Rayleigh-Plateau instability, whereas it opposes the Rayleigh-Taylor instability. 
\citet{Duclaux2006} complemented their experimental measurements with a theoretical analysis based on the lubrication theory and assuming a circular interface. This analysis correctly predicted the scaling of the stability threshold in terms of film thickness and Bond number and qualitatively captured the increasing trend of the instability wavelength with the initial radius of the liquid-air interface. It did not, however, achieve a fully satisfactory quantitative agreement.
In this study, we set to investigate accurately how gravity-induced drainage suppresses the Rayleigh-Plateau instability.
Crucially, we consider the full Navier-Stokes equations and take into account the temporal evolution of the interface.
We study the linear instability in two ways: (i) with a linear stability analysis performed at each instant taken in isolation, assuming that the base flow is "frozen"; (ii) with a transient growth analysis rigorously accounting for the temporal evolution of the base flow.

This paper is structured as follows. The methodology is presented in \S\ref{sec:internaldrainage-Gov-eq}. The problem formulation and the dimensionless governing equations are presented in \S\ref{subsec:internaldrainage-prob-formulation}, from which the base flow is deduced and discussed in \S\ref{subsec:internaldrainage-baseflow}. In \S\ref{subsec:internaldrainage-formulation-stability-analysis}, the stability analysis formulation and the linearised governing equations are elaborated.
Then, in \S\ref{subsec:internaldrainage-formulation-TG} the formulation of the linear transient growth of perturbations is detailed. Corresponding numerical methods are detailed in \S\ref{subsec:internaldrainage_Numerical-method}.
In \S\ref{sec:internaldrainage_results}, the results of the stability and transient growth analyses are presented and discussed.
Section~\ref{subsec:internaldrainage_results_param} summarises the effect of different dimensionless parameters on the stability of the flow:~\S\ref{subsec:internaldrainage_results_Bo}-\ref{subsec:internaldrainage_results_Oh} present the effect of surface tension and inertia, respectively, and~\S\ref{subsec:internaldrainage_results_bifmap} presents the linear stability diagram.
The results of the transient growth analysis are presented in \S\ref{subsec:internaldrainage_results_TG} and finally, conclusions are drawn in \S\ref{sec:internaldrainage_conclusion} by comparing the linear stability and transient growth analyses.
	
\section{Governing equations and methods} \label{sec:internaldrainage-Gov-eq}

\subsection{Problem formulation} \label{subsec:internaldrainage-prob-formulation}

\begin{figure}
	\centerline{\includegraphics[width=0.7\textwidth]{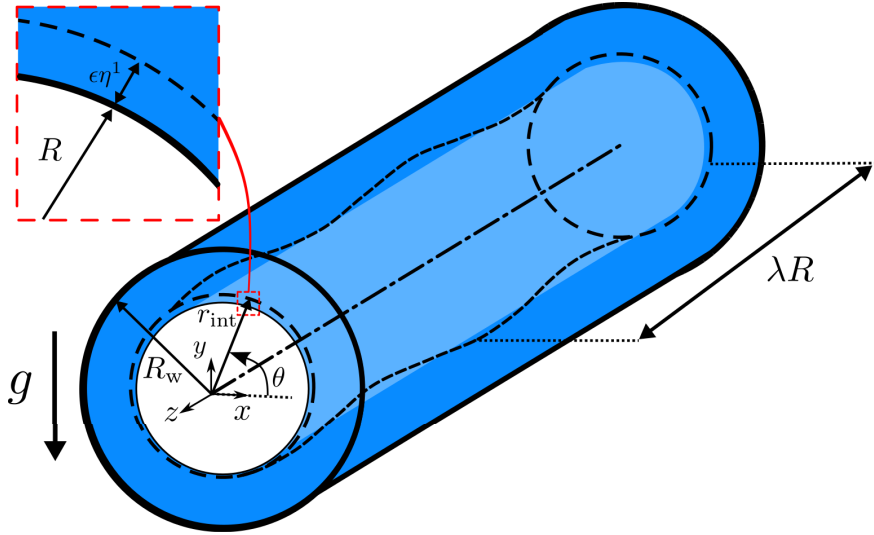}}
	\caption{Schematic of the liquid film coating the inner side of a horizontal tube and the geometrical parameters. The thick solid black line shows the tube wall of inner radius $R_\text{w}$, which is concentric with the coordinate reference. Initially surrounded by a constant film thickness, the liquid-gas interface is a cylinder of radius $R$, concentric with the tube. The dashed black line represents the perturbed liquid-gas interface of local radius $r_\text{int}$ and axial wavelength $\lambda R$. The inset shows the zoomed cross-section of the initial perturbed interface. The gravitational field acts vertically, perpendicular to the tube axis.} 
	\label{fig:internaldrainage_schematic}
\end{figure}

The inner side of a cylindrical tube of radius $R_\text{w}$ is coated with a viscous liquid film. The schematic of the flow is shown in figure~\ref{fig:internaldrainage_schematic}.
The standard Cartesian coordinates $(x,y,z)$ are considered with the origin located at the center of the tube cross-section. In-plane coordinates are $(x,y)$ and the gravity acceleration, ${\bf g}$, points in $-y$ direction.
The liquid is Newtonian, of constant dynamic viscosity $\mu$, surface tension $\gamma$, and density $\rho$, and surrounds a core bubble of inviscid gas of density much smaller than that of the liquid. 
The bubble is initially concentric with the tube and the liquid film thickness is constant on the wall, $h_0=R_\text{w}-R$, where $R$ denotes the initial bubble radius.
The bubble interface can be parametrised without loss of generality in cylindrical coordinates $(r,\theta,z)$ as $r_\text{int}(t,\theta,z)$ using the tube center as the origin. It will be shown in~\S~\ref{subsec:internaldrainage_results_param} that it remains radially representable at later times.
The dimensionless state vector ${\bf q}= ({\bf u},p,\Rb)^T$ defines the flow at time $t$, where~${\bf u}(t,x,y,z)= (u_x,u_y,u_z)^T$ denotes the three-dimensional velocity field, $p(t,x,y,z)$ denotes the pressure, and $\mathcal{R}_\text{int}=r_\text{int} / R$ denotes the dimensionless interface radius.
The state vector and the governing equations are rendered dimensionless by the intrinsic velocity scale associated with a viscous film of thickness $h_0$ falling under its weight, inspired from~\citet{Duclaux2006}. However, we choose differently the length, time, and pressure scales as follows:

\begin{equation} 
	\begin{aligned} \label{eq:internaldrainage-Gauges}
		\mathcal{L}&=R , & \mathcal{U}&=\frac{\rho g h_0^2}{\mu}=\frac{\rho g R^2}{\mu} (\beta-1)^2, \\
		\mathcal{P}& =\frac{\gamma}{R}, &\mathcal{T}&=\frac{\mathcal{L}}{\mathcal{U}}= \frac{\mu}{\rho g R} (\beta-1)^{-2}, 
	\end{aligned}
\end{equation}
where $\beta= {R_\text{w}}/{R}$ denotes the dimensionless tube radius. As a result, the dimensionless value of the initial film thickness can be expressed as $\delta = h_0 / R = \beta -1$.
The flow is governed by the incompressible Navier-Stokes equations which in dimensionless form read
\begin{equation} \label{eq:internaldrainage-Incompressibility}
	\nabla \cdot \,{\bf u}=0,
\end{equation}
\begin{equation} \label{eq:internaldrainage-Navier-Stokes-dimless}
	\left(\frac{Bo}{Oh}\right)^2 \delta^4 \left( \p_t + {{\bf u}}\cdot \nabla \right) {\bf u} =\nabla\cdot\,\stau - Bo \text{ \bf e}_y,
\end{equation}
where $\p_j$ denotes the partial derivative with respect to quantity $j$, and the stress tensor \(\stau\) reads
\begin{equation} \label{eq:internaldrainage-Stresstensor-dimless}
	\stau=-\, {p} \mathsfbi{I} + Bo \ \delta^2 \left( \nabla {{\bf u}} + \nabla {{\bf u}}^T \right). 
\end{equation}
The two other dimensionless numbers which appear in the governing equations are the \textit{Ohnesorge} number, $Oh={\mu}/{\sqrt{\rho \gamma R}}$, and the \textit{Bond} number, $Bo={\rho g R^2}/{\gamma}$.
While $Oh$ compares the viscous forces to the inertial and surface tension forces, $Bo$ compares the gravitational and surface tension forces.

The no-slip boundary condition, $\bf u = 0$, is applied at the tube wall, $r=\beta$. At the shear-free liquid-gas interface, the dimensionless kinematic and dynamic boundary conditions write
\begin{equation} \label{eq:internaldrainage-kinematic-BC}
	\p_t \Rb + \left( {\bf u} \cdot \nabla \right) \Rb= {\bf u \cdot e}_r \quad \text{ at } \ r=\Rb,
\end{equation}
\begin{equation} \label{eq:internaldrainage-dynamic-BC}
	\stau \  {{\bf n}} = \kappa {{\bf n}}  \quad \text{ at } \ r=\Rb,
\end{equation}
respectively, where ${\bf e}_r$ denotes the unit radial vector, ${\bf n} = \nabla \left( r-\Rb\right) / \left\Vert \nabla \left( r-\Rb\right) \right\Vert$ denotes the interface unit normal vector pointing from the gas to the liquid, $\left\Vert \cdot \right\Vert$ denotes the Euclidean norm, and $\kappa=\nabla \cdot {{\bf n}}$ denotes the interface mean curvature.
\subsection{Base flow} \label{subsec:internaldrainage-baseflow}
\begin{figure}
	\centerline{\includegraphics[width=1\textwidth]{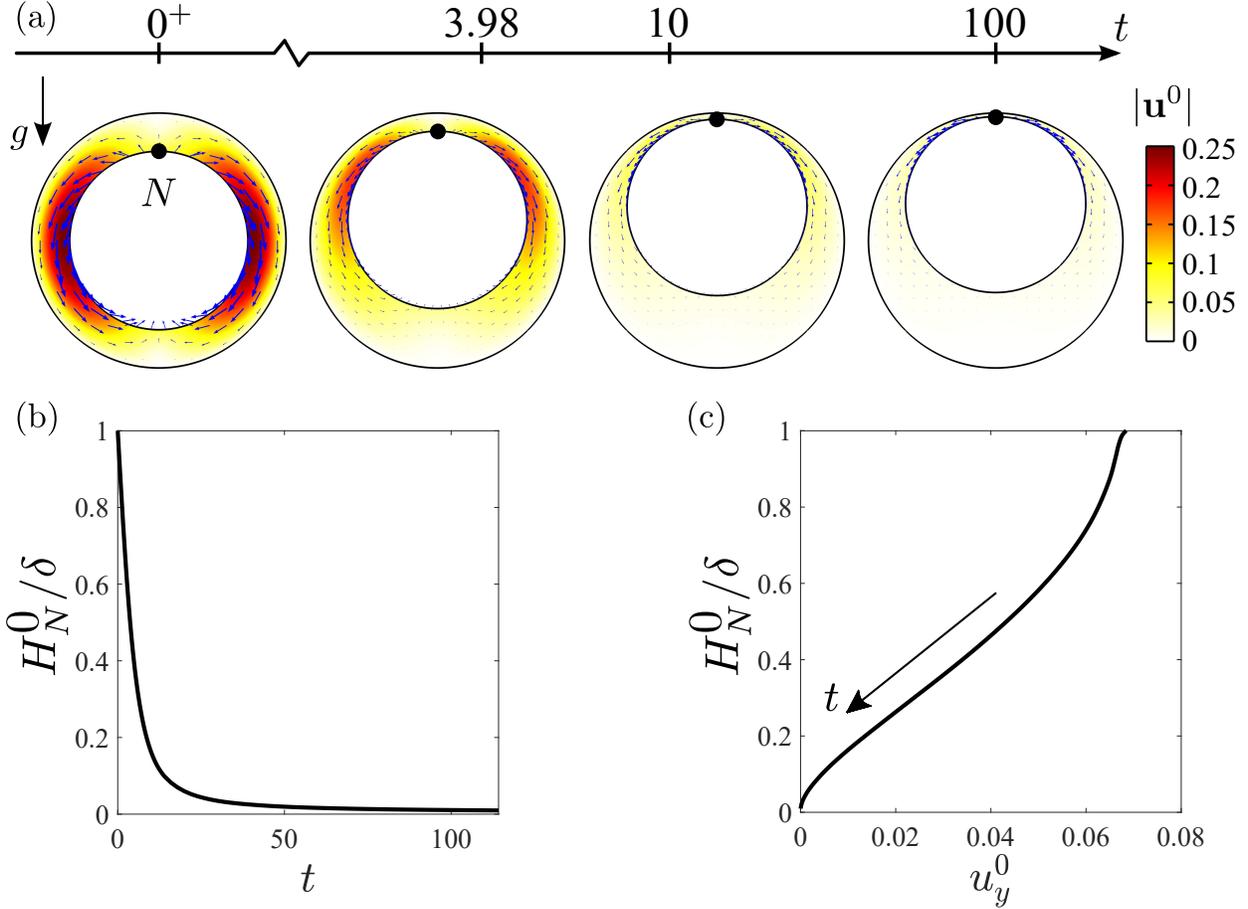}} 
	\caption{Base flow evolution: (a) Snapshots of the flow field: colour map shows the velocity magnitude, arrows show the liquid velocity field, and the point $N$ marks the north pole of the bubble, $\theta=\pi/2$; (b) Temporal variation of the relative liquid film thickness at $N$; (c) Film thickness variation as a function of the vertical velocity at $N$; $Oh \rightarrow \infty, Bo=0.05, \delta / \beta=0.3$. }	\label{fig:internaldrainage-base-flow}
\end{figure}
The {\it base flow} ${\bf q}^0(t,x,y)$ is the two-dimensional time-dependent solution of the Navier-Stokes equations~(\ref{eq:internaldrainage-Incompressibility})-(\ref{eq:internaldrainage-dynamic-BC}) initiated from rest at $t=0$, with pressure $p^0(t=0)=-1 - Bo \ y $, which consists of a constant contribution from the interface Laplace pressure and a linear hydrostatic term.
To begin with, before even presenting the numerical tools in~\S~\ref{subsec:internaldrainage_Numerical-method}, we build up the intuition about the flow
by presenting an illustrative example of the base flow evolution in figure~\ref{fig:internaldrainage-base-flow} for a case where surface tension dominates gravity, $Bo=0.05$, and a thick initial film, $\delta/\beta=0.3$, considering the quasi-inertialess limit when $Oh\rightarrow \infty$.
The liquid drains around the core bubble under gravity, and consequently, the bubble moves upwards under the buoyancy effect.
The dynamics of the base flow, presented in figure~\ref{fig:internaldrainage-base-flow}(b-c), can be characterised by quantifying the liquid film thickness and the vertical velocity at the north pole of the bubble ($N$, hereafter referred to as the north pole), where the strongest gravitational effects are expected~\citep{Trinh2014}.
In the quasi-inertialess limit, $Oh\rightarrow \infty$, drainage begins with an immediate upward drift of the bubble (denoted as $t=0^+$ in figure~\ref{fig:internaldrainage-base-flow}(a)).
Drainage and hence the rising velocity of the bubble decrease with time, as the bubble approaches the top side of the tube.
\subsection{Linear stability analysis} \label{subsec:internaldrainage-formulation-stability-analysis}

Since the base flow presented in \S\ref{subsec:internaldrainage-baseflow} is time-dependent, in order to perform the linear stability analysis, we look at the frozen frames of the unperturbed flow, from which we investigate the evolution of small {\it perturbations} applied to each time frame $t^0$,
sequentially.
Here, we assume that the perturbations evolve much faster than the base flow itself, such that within their evolution, the base flow can be approximated as quasi-steady, an assumption to be verified {\it a posteriori} in \S\ref{subsec:internaldrainage_results_bifmap}. 
To this aim, the state vector ${\bf q}= ({\bf u},p,\Rb)^T$ is decomposed into the sum of the
%
frozen
base flow solution ${\bf q}^0$, and the infinitesimal time-dependent {\it perturbation} ${\bf q}^1= \left( {\bf u}^1,p^1,\eta^1 \right)^T$ which writes 
\begin{equation} \label{eq:internaldrainage-Solution-decomposition}
	{\bf q}={\bf q}^0\vert_{t^0} + \epsilon {\bf q}^1 + \mathcal{O}(\epsilon^2), \quad \epsilon \ll 1,
\end{equation}
where the amplitude $\epsilon$ is assumed to be small. The ansatz of the perturbation ${\bf q^1}$ with the longitudinal wavenumber $k$ (associated with the wavelength $\lambda={2 \pi}/{k}$) reads
\begin{equation} \label{eq:internaldrainage_eigenmode_ansatz}
	{\bf q}^1=\tilde{{\bf q}}(x,y) \ \mathrm{exp} \left[ \sigma t+\mathrm{i}kz \right] + \mathrm{c.c.},
\end{equation} 
where c.c. denotes the complex conjugate. Any other function in terms of the state vector can be decomposed in the same fashion, e.g. $ \stau=  \stau^0+ \epsilon \stau^1 $, $ {\bf n}=  {\bf n}^0+ \epsilon {\bf n}^1 $ and $ \kappa=  \kappa^0+ \epsilon \kappa^1 $. For further details about the formulation of ${\bf n}^1$ and $\kappa^1$, see appendix~\ref{app:internaldrainage_interfacecharacterization}.
In the asymptotic limit of large times, a normal eigenmode perturbation with complex pulsation $\sigma=\sigma_r + \mathrm{i} \sigma_i$, is considered as {\it unstable} if $\sigma_r > 0$, i.e. if $\sigma$ lies in the unstable complex half-plane. An unstable eigenmode grows exponentially in time with the growth rate $\sigma_r$ (unless otherwise noted, the indices $r$ and $i$ denote the real and imaginary parts of a complex number, respectively).
By casting the perturbed state~(\ref{eq:internaldrainage-Solution-decomposition}) into the governing equations~(\ref{eq:internaldrainage-Incompressibility})-(\ref{eq:internaldrainage-Navier-Stokes-dimless}), and keeping the first-order terms, the linearised equations are obtained as
\begin{equation} \label{eq:internaldrainage-linearized-Incompressibility}
	\nabla\cdot\,{\bf u}^1=0,
\end{equation}
\begin{equation} \label{eq:internaldrainage-linearized-momentum}
	\left(\frac{Bo}{Oh}\right)^2 \delta^4 \left( \p_t {\bf u}^1 + \left({\bf u}^0 \cdot \nabla \right) {\bf u}^1 + \left( {\bf u}^1 \cdot \nabla \right) {\bf u}^0  \right) = \nabla \cdot\, \stau^1.
\end{equation}
The no-slip condition implies ${\bf \tilde{u}=0}$ on the solid wall. The interface conditions~(\ref{eq:internaldrainage-kinematic-BC})-(\ref{eq:internaldrainage-dynamic-BC}), applied on the perturbed liquid interface, can be projected radially onto the base interface and ultimately linearised, a process called {\it flattening} (see~(\ref{eq:internaldrainage-flattening}) in appendix~\ref{app:internaldrainage_interfacebc}). The linearised kinematic condition can be expressed as

\begin{equation} \label{eq:internaldrainage_Interface_BC_linearized_kin}
	\sigma \tilde{\eta} + \underbrace{  \left(  -\p_r u^0_r + \frac{\p_r u^0_\theta \ \p_\theta \Rb^0}{\Rb^0} - \frac{u^0_\theta \ \p_\theta \Rb^0 }{\left(\Rb^0\right)^2} \right) \tilde{\eta} + \frac{u^0_\theta  }{\Rb^0} \p_\theta \tilde{\eta} }_{-\mathsfbi{G}^0 \tilde{\eta}} + \ \frac{\p_\theta \Rb^0}{\Rb^0}  \tilde{u}_{\theta} = {\tilde{u}_r } \quad \text{ at} \ r=\Rb^0,
\end{equation}
where $\left( u^0_r, u^0_\theta \right)^T $ and $\left( \tilde{u}_r, \tilde{u}_\theta \right)^T $ denote the velocity vectors of the base state and perturbations, respectively, represented in the cylindrical coordinates.
Introducing an eigenstate vector of the form~(\ref{eq:internaldrainage_eigenmode_ansatz}) into~(\ref{eq:internaldrainage-linearized-Incompressibility})-(\ref{eq:internaldrainage-linearized-momentum}), combined with~(\ref{eq:internaldrainage_Interface_BC_linearized_kin}), leads to a generalised eigenvalue problem for $\sigma$ and $\tilde{{\bf q}}$ that writes

\begin{equation} \label{eq:internaldrainage-eigenvalue-problem}
	\mathsfbi{L} \tilde{{\bf q}} + \text{c.c.} =\sigma \mathsfbi{B} \tilde{{\bf q}} + \text{c.c.},
\end{equation}
where the linear operators $\mathsfbi{L}$ and $\mathsfbi{B}$ are defined as
\begin{align} \label{eq:internaldrainage_eigenvalue_operators_global}
	\nonumber \mathsfbi{L} & = \begin{bmatrix}
		\left(\frac{Bo}{Oh}\right)^2 \delta^4 \mathsfbi{F}^0 + Bo \delta^2 \left( \tilde{\nabla}\cdot(\tilde{\nabla}+\tilde{\nabla}^T) \right)       & -\tilde{\nabla} & {\bf 0} \\
		\tilde{\nabla}\cdot       & 0 & 0 \\
		\left( {\bf e}_r - \frac{\p_\theta \Rb^0}{\Rb^0} {\bf e}_\theta \right) \cdot       & 0 & \mathsfbi{G}^0
	\end{bmatrix}, \\  
	\mathsfbi{B}& = \begin{bmatrix}
		\left(	\frac{Bo}{Oh} \right)^2 \delta^4 \mathsfbi{I}   & {\bf 0} & {\bf 0}  \\
		{\bf 0} & 0 & 0 \\
		{\bf 0} & 0 & 1 \,  
	\end{bmatrix},
\end{align}
where $\mathsfbi{F}^0 \tilde{\bf u} = -\bigl( \left({\bf u}^0 \cdot \tilde{\nabla} \right) \tilde{\bf u} + \left( \tilde{\bf u} \cdot \nabla \right) {\bf u}^0 \bigr) $, $({ \bf e}_r,{ \bf e}_{\theta},{ \bf e}_z)$ denote the vectors of unit directions in the cylindrical coordinates $(r,\theta,z)$ used for parameterising the interface, and the gradient operators and the velocity gradient tensors in the Cartesian coordinates read

\begin{align} \label{eq:internaldrainage-gradient-operator-global}
	\nonumber	{\nabla} & = (\p_x, \p_y,\p_z)^T, &{\nabla} {\bf u}^0 &= \begin{bmatrix}
		\p_x {u}^0_x & \p_y {u}^0_x & 0  \\
		\p_x {u}^0_{y} & \p_y {u}^0_{y} & 0  \\
		0 & 0 & 0\, 
	\end{bmatrix}, \\
	\tilde{\nabla} & = (\p_x, \p_y,\mathrm{i}k)^T,
	&\tilde{\nabla} \tilde{{\bf u}} &= \begin{bmatrix}
		\p_x \tilde{u}_x & \p_y \tilde{u}_x & \mathrm{i}k \tilde{u}_x  \\
		\p_x \tilde{u}_{y} & \p_y \tilde{u}_{y} & \mathrm{i}k \tilde{u}_{y}  \\
		\p_x \tilde{u}_z & \p_y \tilde{u}_z & \mathrm{i}k \tilde{u}_z  \, 
	\end{bmatrix}. 	\\
\end{align}
The interface dynamic condition~(\ref{eq:internaldrainage-dynamic-BC}), once linearised, can be expressed as
\begin{equation} \label{eq:internaldrainage_Interface_BC_linearized_dyn}
	\stau^0 \tilde{\bf n} + \tilde{\eta} \, \p_r \stau^0 {\bf{n}}^0 + \underline{\underline{{\tilde{\tau}}}} \ {\textbf{n}}^0  =  \kappa^0 \tilde{\bf n} + \tilde{\kappa} {\textbf{n}}^0   \quad \text{ at} \ r=\Rb^0.
\end{equation}
For further details on the derivation of the interface conditions and their implementation, see appendices~\ref{app:internaldrainage_interfacebc} and~\ref{app:internaldrainage_implementation_LSA}, respectively. 

\subsection{Transient growth analysis} \label{subsec:internaldrainage-formulation-TG}

In contrast to the linear stability analysis, the \textit{transient growth analysis} tolerates any kind of temporal dependency for both the base flow and the perturbation and does not rely on a separation of time scales between their respective evolutions. Moreover, it accounts for the so-called {\it "nonmodal"} mechanisms arising from the nonnormality (non-commutativity with the adjoint) of the linear operator $\mathsfbi{L}$. Owing to these mechanisms, small-amplitude initial perturbations may experience a large transient amplification due to an intricate cooperation between a possibly large number of eigenmodes; therefore, reducing the dynamics to the leading (least stable or most unstable) eigenmode might be irrelevant at finite time. If nonnormality is often inherited from the linearization of the advective term, the operator $\mathsfbi{L}$ expressed in (\ref{eq:internaldrainage_eigenvalue_operators_global}) can be nonnormal even in the quasi-inertialess limit $Oh \rightarrow \infty$, due to the presence of the interface. 

The ansatz (\ref{eq:internaldrainage_eigenmode_ansatz}) is generalised as
\begin{equation} 
	{\bf q}^1=\hq(t,x,y) \ \mathrm{exp} \left[\mathrm{i}kz \right] + \mathrm{c.c.},
	\label{eq:internaldrainage_tg_ansatz}
\end{equation} 
where an exponential temporal behavior is not enforced. The temporal evolution of $\hq(t,x,y)$ is an extension of (\ref{eq:internaldrainage-eigenvalue-problem})
\begin{equation} \label{eq:internaldrainage-tg-ode}
	\mathsfbi{L} \hq + \text{c.c.} = \mathsfbi{B} \frac{\partial \hq}{\partial t} + \text{c.c.},
\end{equation}
We recall that $\mathsfbi{L}$ is time-dependent and parameterised by $k$. We are interested in the initial perturbation of the interface $\hq(0,x,y)=[{\bf 0},0,\bn(0)]^T$ that is the most amplified by (\ref{eq:internaldrainage-tg-ode}) after a temporal horizon $t=T$, where $T$ is named a \textit{temporal horizon}. A given initial condition might project on the optimal one, and can be amplified strongly enough to lead to a nonlinear regime. We follow the methodology proposed in~\citet{DelGuercio2014}, and take advantage of the fact that the base flow interface is axisymmetric at $t=0$ (see figure \ref{fig:internaldrainage-base-flow}) to expand $\bn(0)$ as a finite series of Fourier modes in $\theta$
\begin{equation} \label{eq:n0}
\bn(0) = \sum_{m=-N}^{N} \al_m e^{\mathrm{i}m\theta} = a_0 + \sum_{m=1}^{N} [a_m \cos(m\theta) + b_m \sin(m\theta)]
\end{equation}
with 
\begin{equation*}
\alpha_{-m} = \al_m^{\star}
\end{equation*}
where the superscript $\star$ denotes the complex conjugate, $a_0=\alpha_0$, and for $m\geq 1$ 
\begin{equation*}
a_m = \al_m+\al_m^{\star}, \quad \text{and} \quad b_m = i(\al_m-\al_m^{\star}).
\end{equation*}
Note that $a_m$, $b_m$ ($m=0,1,2,...$) and $\bn(0)$ are real-valued. The associated first-order interfacial energy density per spanwise wavelength is proportional to 
\begin{equation}
    e_0 = \frac{k}{2 \pi} \int_{0}^{2\pi} \int_{0}^{2\pi/k}   \left[ \bn(0) e^{\mathrm{i}kz} + c.c \right]^2  \mathrm{d}z \mathrm{d}\theta = 2\pi {\bm a}^T\bm{a},
\end{equation}
where $\Rb^{0}(0)=1$ was implied, and where we defined
\begin{equation}
    \bm{a} = [a_N,a_{N-1},...,a_1,\sqrt{2}a_0,b_N,b_{N-1},...,b_1].
\end{equation}
Let us define $\bn_m(t)$ as the evolution at time $t$ of each term $\bn_m(0)=e^{\mathrm{i} m \theta}$ of the inital condition (\ref{eq:n0}). Thanks to the linearity of the evolution equation (\ref{eq:internaldrainage-tg-ode}), the interface shape at $t=T$ reads simply 
\begin{equation} 
\bn(T) = \sum_{m=-N}^{N} \al_m \bn_m(T).
\end{equation}
The corresponding interfacial energy density per wavelength is proportional to
\begin{equation}
e(T) = \frac{k}{2 \pi} \int_{0}^{2\pi} \int_{0}^{2\pi/k} |\bn(T)e^{\mathrm{i}kz} + c.c |^2 \mathrm{d}z \Rb^{0}(T)\mathrm{d}\theta.
\end{equation}
We further show in appendix~\ref{app:tgmatrix} that $e(T) = 2\bm{a}^T\mE(T)\bm{a}$ where $\mE(T)$ is a real-valued, symmetric, strictly positive definite matrix of size $(2N+1)\times(2N+1)$. In this manner, the (optimal) \textit{transient gain} $G(T)$ defined as
\begin{equation}
G(T) = \max_{\bm{a}}\frac{e(T)}{e_0} = \pi^{-1}\frac{\bm{a}^T\mE(T)\bm{a}}{\bm{a}^T\bm{a}}
\label{eq:tgdef}
\end{equation}
is simply the largest eigenvalue of $\mE(T)$ divided by $\pi$, and the associated eigenvector provides directly the Fourier mode coefficients of the optimal initial condition. 

\subsection{Numerical method} \label{subsec:internaldrainage_Numerical-method}
\begin{figure}	\centerline{\includegraphics[width=0.3\textwidth]{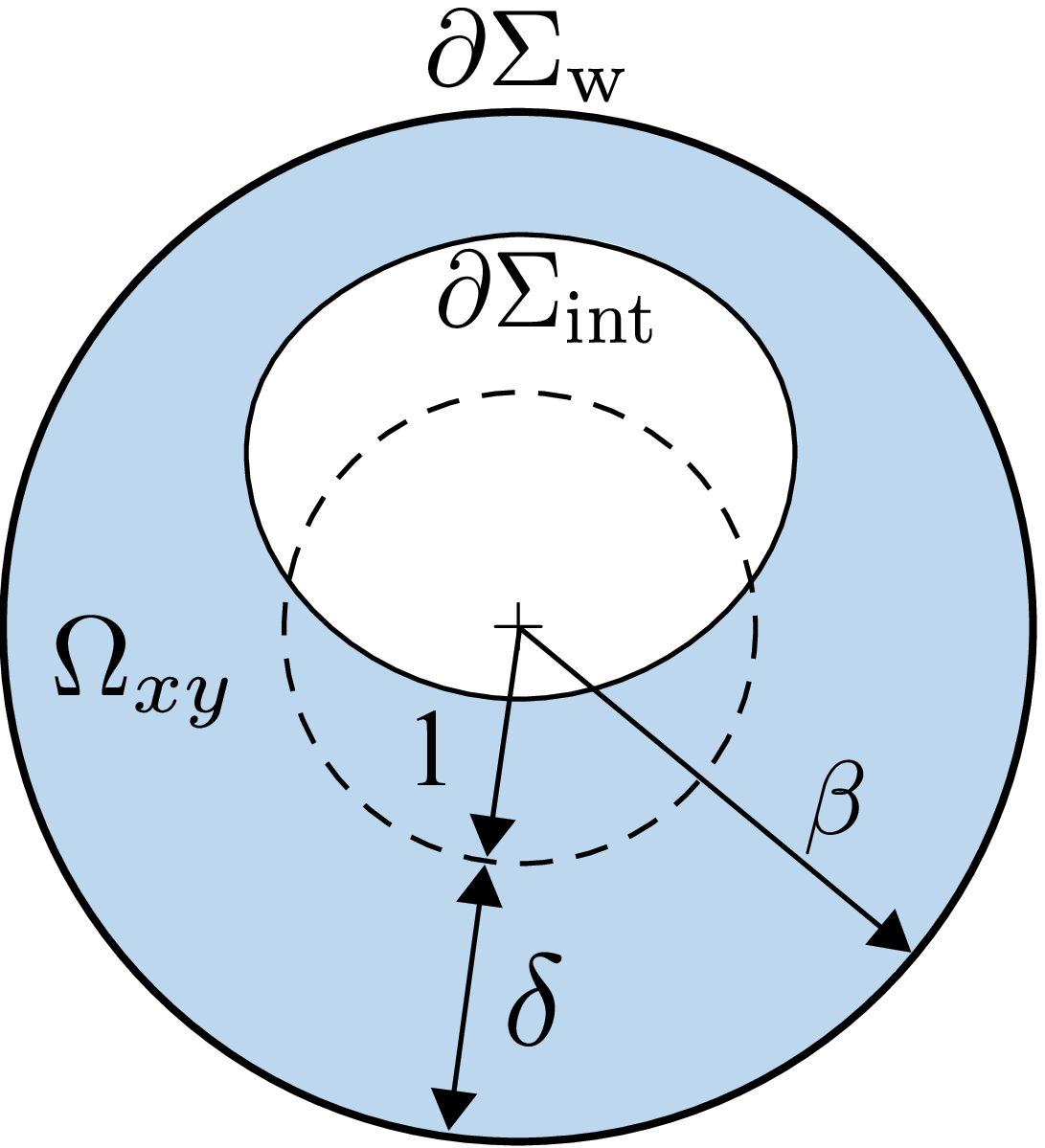}	}
	\caption{\label{fig:internaldrainage_numerical_domain_drainage_cartesian} The numerical domain used for computing the base flow and the linear stability analysis. $\Omega_{xy}$ denotes the liquid bulk. The boundaries of the numerical domain are denoted by $\p \Omega_{xy} = \p \Sigma_{\mathrm{w}} \cup \p \Sigma_{\mathrm{int}}$: $\p \Sigma_{\mathrm{w}}$ represents the interior wall of the tube with the radius of $\beta$, and $\p \Sigma_{\mathrm{int}}$ represents the gas-liquid interface. The cross-section of the interface is initially a circle of unit radius, concentric with the tube (sketched with the dashed line).}
\end{figure}
\indent The base flow calculation, linear stability, and transient growth analyses are performed numerically employing the finite element package COMSOL Multiphysics$^\text{TM}$. 
A triangular mesh is generated on the two-dimensional domain shown in figure~\ref{fig:internaldrainage_numerical_domain_drainage_cartesian}. In the following sections, the area increment in the bulk cross-section is denoted by $ \mathrm{d} A_{\Omega_{xy}}$. On the boundary $j$, the increment of the surface area is denoted by $ \mathrm{d} A_{\Sigma_j}$, and the increment of arc length is denoted by $ \mathrm{d} s$. The grid size is controlled by the vertex densities on the boundaries $\p \Sigma_{\mathrm{w}}$ and $\p  \Sigma_{\mathrm{int}}$.
The variational formulations of the base flow equations~(\ref{eq:internaldrainage-Incompressibility})-(\ref{eq:internaldrainage-dynamic-BC}), linear stability equations (\ref{eq:internaldrainage-eigenvalue-problem}), and linearised Navier-Stokes equations (\ref{eq:internaldrainage-tg-ode}) are discretised spatially using quadratic (P2) Lagrange elements for ${\bf u}^0$, $\tilde{{\bf u}}$, ${\bf \bar{u} }$, $\tilde{\eta}$ and $\bn$ as well as the base flow interface geometry, and linear (P1) Lagrange elements for $p^0$, $\tilde{p}$, and $\bar{p}$.
While the interface conditions are applied in the built-in modules of COMSOL Multiphysics$^\text{TM}$, the linearised condition (\ref{eq:internaldrainage_Interface_BC_linearized_dyn}) is applied by means of the Lagrange multipliers of quadratic (P2) order.
The employed discretisation yields approximately 500'000 degrees of freedom for the base flow, linear stability, and transient growth analyses.

\indent The time-dependent base flow is computed using the laminar two-phase flow module coupled with the moving mesh module. 
The numerical time step is set by the Backward differentiation formula with maximum differentiation order of 2.
The solver is initialised by the Backward Euler consistent initialization with an initial step fraction of $10^{-9}$. 
At each time step, Newton's method is used to solve the non-linear equations, where the relative tolerance for convergence is set to $10^{-6}$.
Following the computed base flow, and after extracting the geometric characteristics of the base interface at each time step, the generalised eigenvalue problem~(\ref{eq:internaldrainage-eigenvalue-problem}) is solved using the shift-invert Arnoldi method (for more details on the development of the variational formulation, the implementation of the linear stability eigenvalue problem and their corresponding boundary conditions see appendix~\ref{app:internaldrainage_implementation_LSA}). 

\indent The computation time associated with obtaining the base flow for a given set of parameters, followed by the stability analysis for~$\sim$20 values of $k$, is of the order of a few hours on a single Intel core at 3.6 GHz.
For the transient growth analysis, the computation of the propagator matrices with $-5  \le m  \le 5$ over 10 values of $k$ takes a few days; the calculation of the optimal transient gain for the same $k$ values and~$\sim$12 values of the temporal horizon $T$ takes approximately one hour.
Both the base flow and stability analysis model are validated with the analytical solutions available in the literature for the flow of liquid coating the inside of a vertical tube (for more details about the series of validation tests see appendix~\ref{app:internaldrainage_validation}).
\section{Results} \label{sec:internaldrainage_results}
\subsection{Linear stability analysis} \label{subsec:internaldrainage_results_param}
In this section, we present the linear stability characteristics associated with the drainage of a liquid film coating the inside of a horizontal circular tube.
The results of the frozen base flow linear stability analysis in terms of different parameters are presented hereafter.
First, an overview of the stability characteristics of the flow is presented in \S\ref{subsec:internaldrainage_results_LSA_overview}.
Then, the influence of $Bo$ and $Oh$ is demonstrated in \S\ref{subsec:internaldrainage_results_Bo}-\ref{subsec:internaldrainage_results_Oh}, and the stability diagram is presented in \S\ref{subsec:internaldrainage_results_bifmap}.
Additional comments on the effect of $Bo$ and on the most unstable wavenumber are given in \S\ref{subsec:internaldrainage_results_bondisolation}-\ref{subsec:internaldrainage_results_kmax}.
The validity of the frozen base flow assumption is discussed in \S\ref{subsec:internaldrainage_results_frozendic}.
\subsubsection{Stability of draining film at different instants of its evolution} \label{subsec:internaldrainage_results_LSA_overview}
\begin{figure}
	\centerline{\includegraphics[width=1\textwidth]{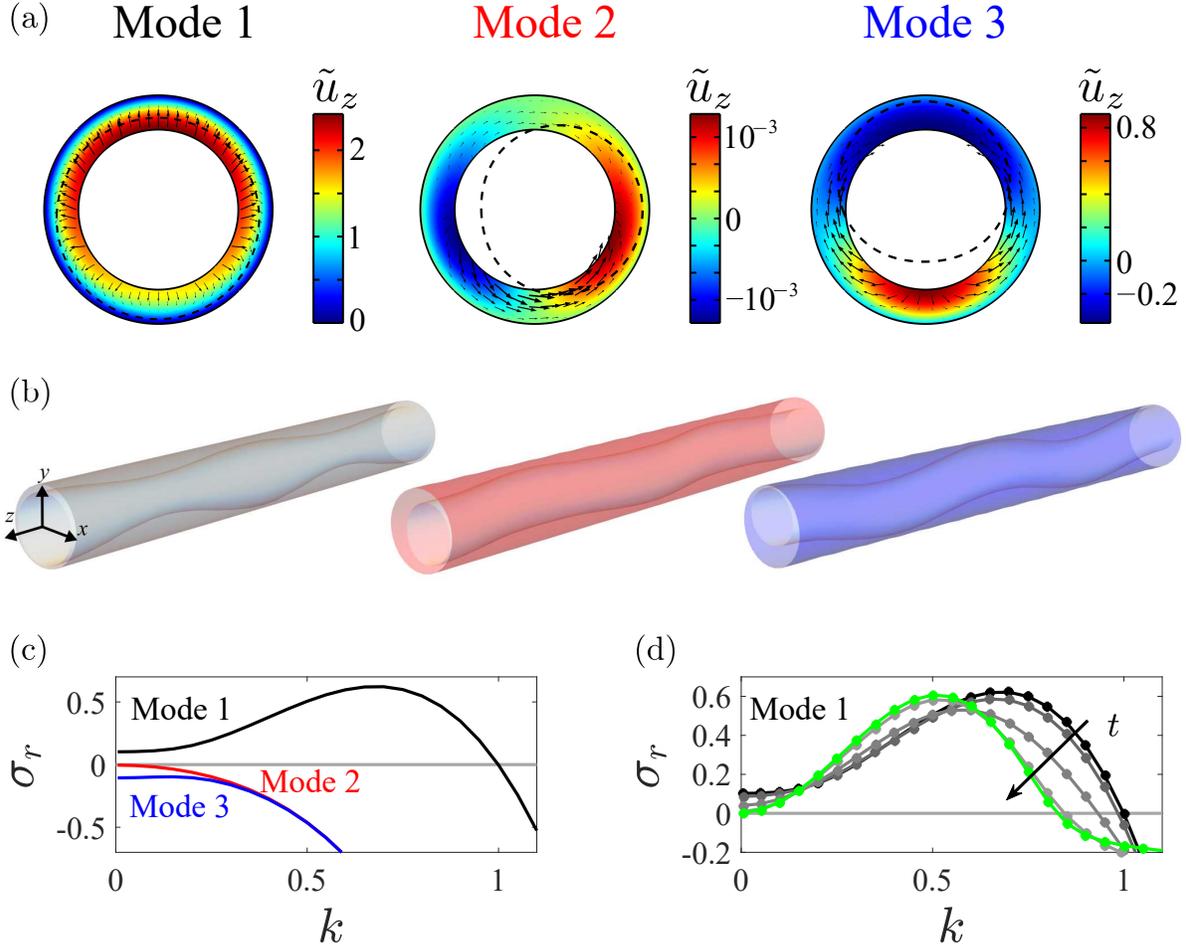}} 
	\caption{Linear stability analysis of the base flow presented in figure~\ref{fig:internaldrainage-base-flow}; (a) Three least stable eigenmodes for $t = 0^+, k=0.55$; colour map presents the axial eigenvelocity, black arrows show the in-plane eigenvelocity field, and black dashed line shows the superposition of the eigeninterface (with arbitrary amplitude) onto the base interface;
    (b) Three-dimensional render of the same perturbed interfaces;
    (c) The dispersion curve of the same three modes at $t=0^+$;
    (d) Temporal variation of the dispersion curve of the unstable mode (mode 1); The green line marks the large time limit; $Oh \rightarrow \infty, \ \delta / \beta=0.3 , \ Bo=0.05$.}
	\label{fig:internaldrainage_LSA_overview}
\end{figure}
Figure~\ref{fig:internaldrainage_LSA_overview}(a-b) shows the three least stable modes of the base flow presented in figure~\ref{fig:internaldrainage-base-flow} at the initial time, $t=0^+$, and figure~\ref{fig:internaldrainage_LSA_overview}(c) presents the corresponding {\it dispersion curves} (growth rate $\sigma_r$ as a function of the axial wavenumber $k$).
Only one unstable mode (mode 1) is detected, whose dispersion curves at later time instants are shown in figure~\ref{fig:internaldrainage_LSA_overview}(d).
The characteristics of this mode are symmetry with respect to the vertical mid-plane, strong interface modulation at the bottom of the bubble, and weak interface modulation at the top of the bubble.
The fact that this mode is unstable at $k=0$ is reminiscent of the instability properties of a purely viscous liquid thread in absence of inertia (see \citet{Eggers2008}).
This mode resembles the structure of the unstable interface observed in the experiments of~\citet{Duclaux2006}.
The maximal growth rate of the unstable mode initially decreases before increasing and eventually saturating to a value inferior to the initial one (see the green line in figure~\ref{fig:internaldrainage_LSA_overview}(d)), as the flow evolves further. The corresponding wavenumber decreases monotonously.

\subsubsection{Effect of the Bond ($Bo$) number} \label{subsec:internaldrainage_results_Bo}
\begin{figure}
	\centerline{\includegraphics[width=1\textwidth]{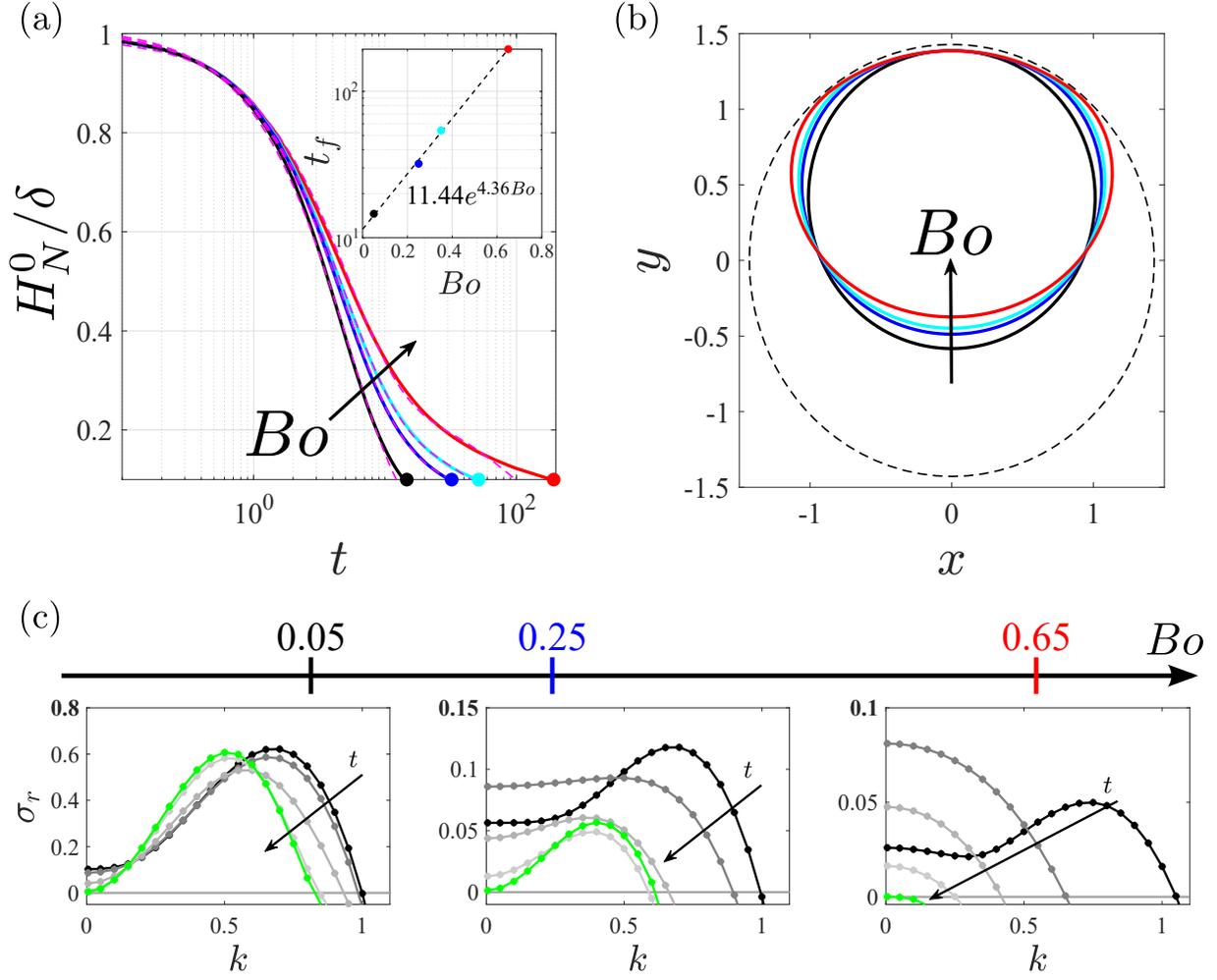}} 
	\caption{Influence of $Bo$ on the base flow and its long-time linear stability for $Oh \rightarrow \infty, \ \delta / \beta=0.3 , \ Bo=\{0.05,0.25,0.35,0.65\}$;
    (a) Evolution of the relative thickness at the north pole; 
    dashed line present the exponential fit $H_N^0/\delta = a_1 \exp[-t/T_1] + a_2 \exp[-t/T_2]$ and the values of $\{a_1,a_2,T_1,T_2\}$ are given in table~\ref{tab:internaldrainage_Bo_effect_BF_fit};
    inset shows the variation of $t_f$ as a function of $Bo$;
    (b) Bubble interface at $t=t_f$ when the relative thickness at the north pole reaches $10 \%$; dashed line shows the tube wall;
    (c) Dispersion curve of the unstable eigenmode, mode 1, for $Bo=\{0.05,0.25,0.65\}$.}
	\label{fig:internaldrainage_Bo_BF}
\end{figure}
\begin{table}
	\begin{center}
		\def~{\hphantom{0}}
		\begin{tabular}{ccccc}
			$Bo$ 		   & $a_1$ 		& $T_1$      & $a_2$ 	   & $T_2 $   \\ [10pt]
			0.05           & 0.986      & 5.07      & 0.014       & 51.5        \\
			0.25           & 0.831      & 4.75      & 0.178       & 52.9		\\
			0.35           & 0.788      & 4.74      & 0.217       & 58.0		\\
			0.65           & 0.731      & 5.07      & 0.261       & 103
		\end{tabular}
		\caption{Fitting parameters associated with the relative pole thickness $H_N^0/\delta = a_1 \exp[-t/T_1] + a_2 \exp[-t/T_2]$ for different values of $Bo$ presented in figure~\ref{fig:internaldrainage_Bo_BF}(a).}
		\label{tab:internaldrainage_Bo_effect_BF_fit}
	\end{center}
\end{table}
In this section, the influence of $Bo$ on the base flow and its stability is illustrated.
Figure~\ref{fig:internaldrainage_Bo_BF}(a-b) show for several $Bo$ values the characteristics of the north pole evolution until $t=t_f$ when the relative thickness at the north pole, $H_N^0/\delta$, diminishes below $10 \%$ for an initial thick film $\delta/\beta=0.3$.
The relative pole thickness decays exponentially with two distinct time scales, presented in table~\ref{tab:internaldrainage_Bo_effect_BF_fit}.
Increasing $Bo$ results in an exponential increase of $t_f$ (figure~\ref{fig:internaldrainage_Bo_BF}(a)).
Figure~\ref{fig:internaldrainage_Bo_BF}(b) presents for the same flows the bubble interface at $t=t_f$. This figure evidences the horizontal bubble widening as it rises towards the upper wall. Low surface tension or large gravitational effects, i.e. high $Bo$, results in more deviation of the interface cross-section from its initial circular shape.
Note that as $Bo$ increases, the location where the gap between the interface and the tube is thinnest moves from the north pole to the sides.

The influence of $Bo$ on the dispersion curve of mode 1 is presented in figure~\ref{fig:internaldrainage_Bo_BF}(d).
The evolution of the dispersion curves for small $Bo$ values (here $Bo=\{0.05,0.25\}$) is similar to that demonstrated in \S\ref{subsec:internaldrainage_results_LSA_overview}.
Increasing $Bo$ reduces the maximal growth rate of the unstable mode and the range of unstable wavenumbers, $k$. Further increasing $Bo$ results in the suppression of the instability after a finite time (see $Bo=0.65$). In this investigation, no other unstable or marginally stable mode was detected while varying $Bo$.
\begin{figure}
	\centerline{\includegraphics[width=1\textwidth]{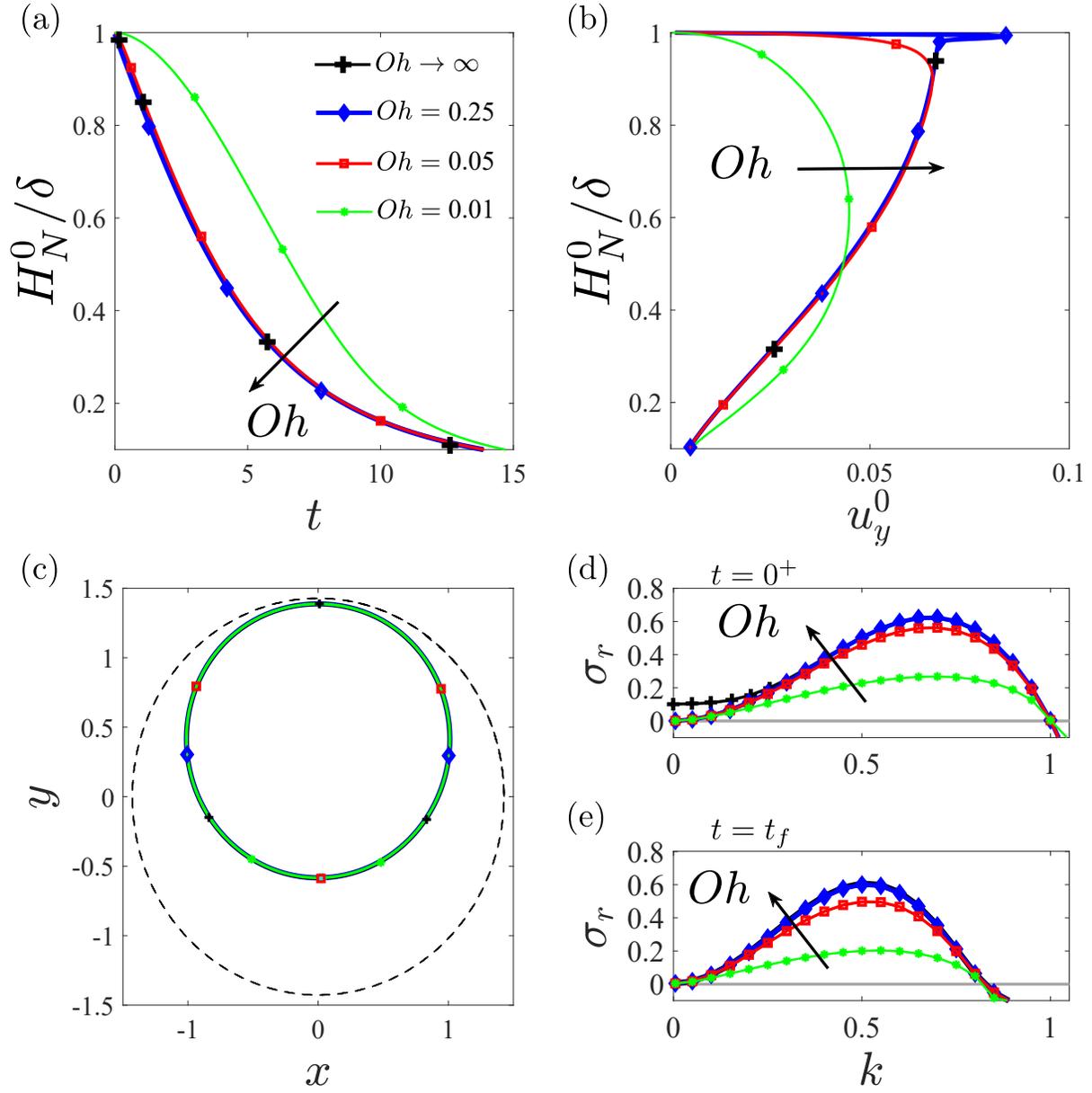}} 
	\caption{Influence of $Oh$; (a) Evolution of the relative thickness at the north pole;
    (b) The relative film thickness variation as a function of the velocity at the north pole;
    (c) Interface at $t=t_f$ when the relative thickness at the north pole reaches $10 \%$; dashed line ($--$) shows the tube wall;
    (d) Dispersion curve at $t=0^+$;
    (e) Dispersion curve at $t=t_f$; $Bo=0.05, \ \delta / \beta=0.3 , \ Oh=\{0.01,0.05,0.25,\infty\}$ (corresponding to $0 \le ({Bo}/{Oh})^2 \delta^4 \le \mathcal{O}(1)$).}
	\label{fig:internaldrainage_Oh}
\end{figure}

\clearpage
\newpage

\subsubsection{Effect of the Ohnesorge ($Oh$) number}\FloatBarrier \label{subsec:internaldrainage_results_Oh}
In the present study, all of the aforementioned results were limited to the quasi-inertialess regime, where $	\left({Bo}/{Oh}\right)^2 \delta^4 \rightarrow 0$.
One can interpret $\left({Bo}/{Oh}\right)^2 \delta^4 = Re \, \delta^2 $ where the Reynolds number, $Re$, is defined upon the same length scale and drainage velocity scale given in (\ref{eq:internaldrainage-Gauges}).
In this section, we study the base flow and its stability in presence of inertia by decreasing $Oh$ while keeping the rest of the flow parameters identical to those presented in \S\ref{subsec:internaldrainage-baseflow} and \S\ref{subsec:internaldrainage_results_LSA_overview}.
Figure~\ref{fig:internaldrainage_Oh}(a-b) depict the influence of inertia on the base flow evolution from the initial time $t=0^+$ to a terminal time $t=t_f$ identical to that presented in \S\ref{subsec:internaldrainage_results_Bo}.
Compared to the quasi-inertialess drainage, the inertial bubble accelerates and decelerates with some delay.
Eventually, the relative film thickness at the north pole and its velocity converge to that of the quasi-inertialess flow. The delays in the bubble acceleration and deceleration increase by decreasing $Oh$. However, the bubble interfaces in all cases are identical at the terminal time (figure~\ref{fig:internaldrainage_Oh}(c)).

Figure~\ref{fig:internaldrainage_Oh}(d-e) present the effect of $Oh$ on the dispersion curve of the unstable mode at $t=0^+$ and $t=t_f$, respectively.
Including inertia (finite $Oh$) affects the initial dispersion relation of the unstable mode in two ways. The first effect is associated with $k=0$, where the immediate bubble drift in the inertialess flow was seen to induce a non-zero growth rate.
The addition of inertia, no matter small, yields a zero growth rate, a regularisation by inertia also observed for the Rayleigh-Plateau instability of a liquid thread (see~\citet{Eggers2008}).
The second effect is linked to the shape of the dispersion curve, which is similar to that of the classical thin-film Rayleigh-Plateau instability in the vicinity of a wall and in the absence of gravity, as obtained by a lubrication analysis $\sigma_r \propto k^2-k^4$~\citep{Camassa2014,Gallaire2017}, for $Oh>0.25$.
In the limit of small $Oh$, inertia diminishes the growth rate of the perturbations without altering the instability cut-off compared to the inertialess limit. 
\begin{figure}
	\centerline{\includegraphics[width=1\textwidth]{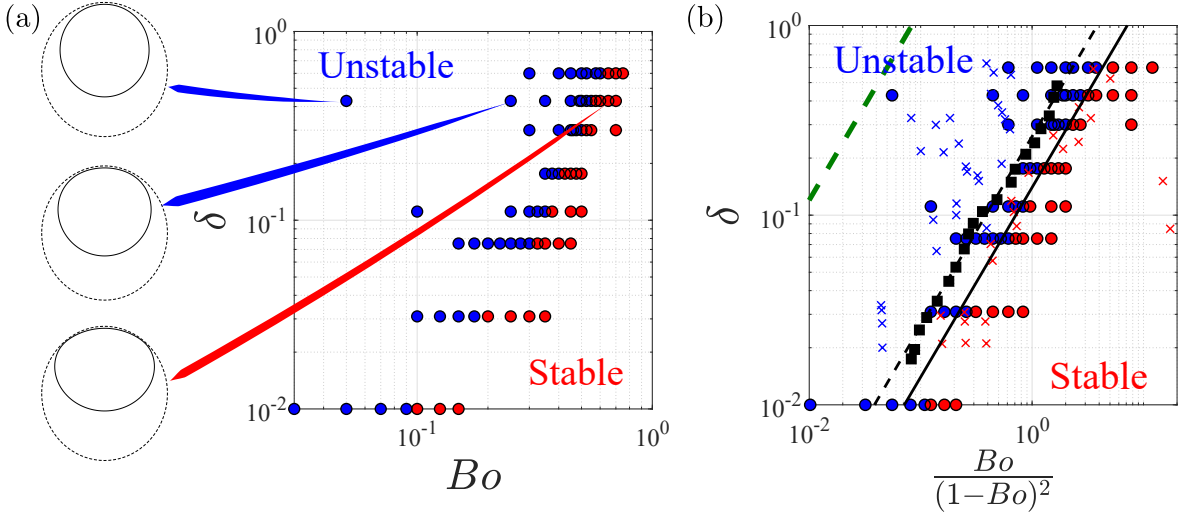}} 
	\caption{Linear stability stability diagram; (a) in $\delta$ vs $Bo$ plane; the blue and red bullets are obtained from the present numerical stability analysis;
    three base interfaces at $t=t_f$ are shown for $\delta / \beta = 0.3$ (corresponding to $\delta=0.43$) and $Bo =\{ 0.05, 0.25,0.65\};$ 
    (b) the same diagram shown in $\delta$ vs $Bo / ( 1-Bo )^2$ plane; 
    the green dashed line shows the analytical stability margin predicted by~\citet{Duclaux2006}, the black squares and the passing dashed line show their transition experimental data and their best linear fit, $\delta = 0.26 Bo / (1-Bo)^2$, respectively; 
    the crosses show their experimental data; the continuous black line shows the best linear fit to the present linear analysis: $\delta = 0.14 Bo / (1-Bo)^2$;
}	\label{fig:internaldrainage_phasediagram}
\end{figure}
\subsubsection{Asymptotic linear stability diagram} \label{subsec:internaldrainage_results_bifmap}
The $\{\delta, Bo\}$ space is investigated to follow the transition of the interface from unstable to stable.
Figure~\ref{fig:internaldrainage_phasediagram}(a) presents the {\it stability diagram} based on the dispersion curve of the most unstable mode obtained from the linear stability analysis at large time $t=t_f$ (it is verified that evaluation of the stability at $t=\{0.75 t_f, 7 t_f\}$ does not alter the obtained marginal curve). For each $\delta$, increasing $Bo$ results in interface stabilisation.
Figure~\ref{fig:internaldrainage_phasediagram}(b) presents the same linear stability phase diagram in $\delta$ versus $Bo / ( 1-Bo )^2$ plane. Such an abscissa is suggested by~\citet{Duclaux2006} who proposed an analytical dispersion relation using the lubrication theory and assuming a circular base interface.
Note that their definition of $Bo$ is different from ours and the abscissa is adapted to the present study.
Even though their analysis suggests the stability threshold as $\delta= 12 Bo/( 1-Bo)^2$, their experiments suggest $\delta= 0.26 Bo / ( 1-Bo )^2$ for the stability transition. 
The present numerical stability analysis suggests the best fit to the separatrix as $\delta= 0.14 Bo / ( 1-Bo )^2$. Therefore, our stability analysis yields a significant improvement, although a discrepancy remains. 
Note that the definition by \citet{Duclaux2006} of the transition regime is not the appearance of growing perturbations but rather that of a visible top/down asymmetry. It is therefore reassuring to observe their transition points (black squares) fall in the unstable region predicted by our stability analysis (see figure~\ref{fig:internaldrainage_phasediagram}(b)).
However, for a few points, the linear analysis predicts instability while the experiments report a stable interface, resulting in a small relative error in $\delta$.
\subsubsection{Why does increasing Bond number stabilise the flow?}\FloatBarrier \label{subsec:internaldrainage_results_bondisolation}
\begin{figure}
	\centerline{\includegraphics[width=1\textwidth]{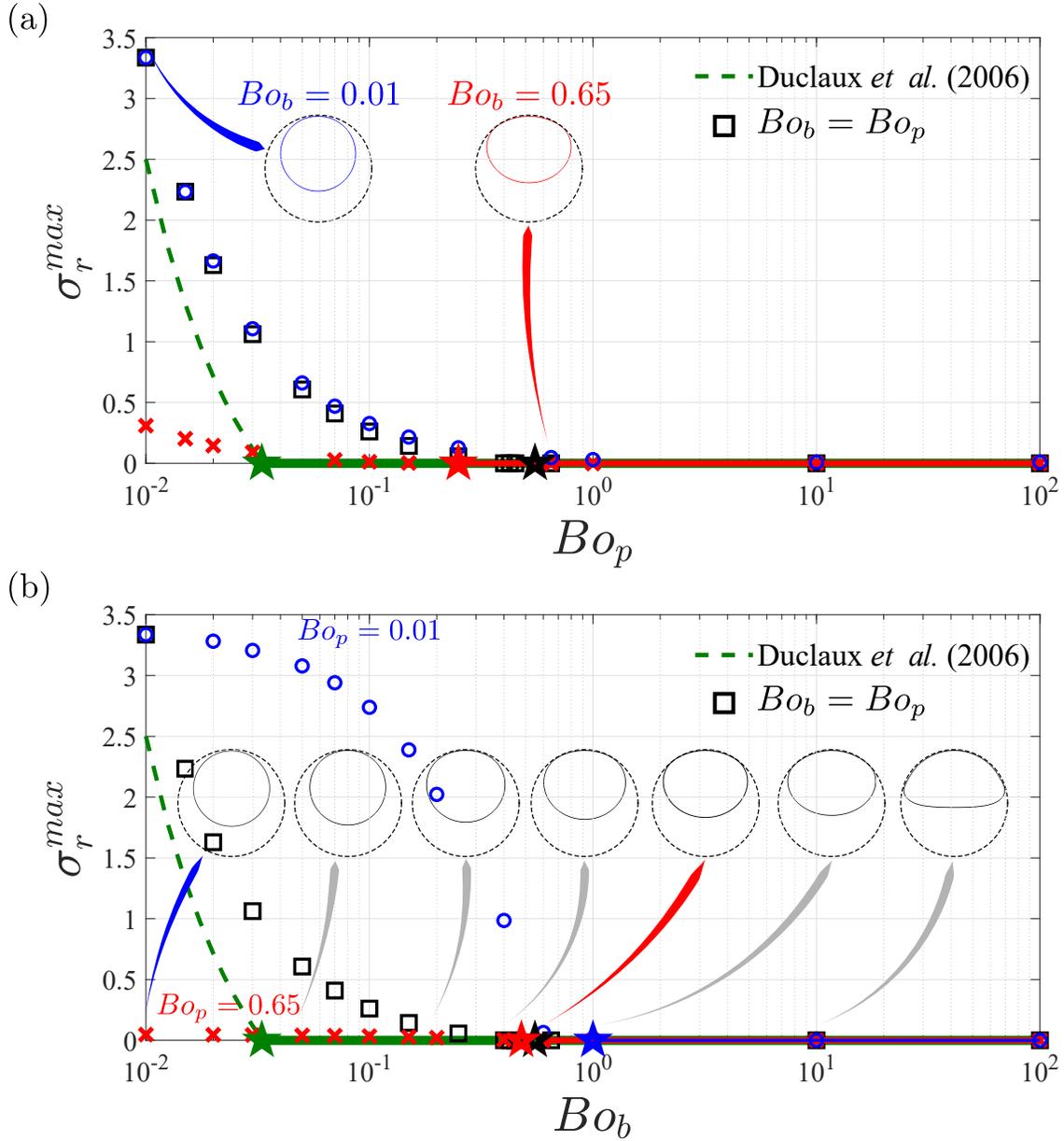}} 
	\caption{Decoupled effects of $Bo$ on the linear stability of the flow at different scales for $\delta / \beta =0.3$:
    (a) the maximal growth rate obtained by isolating $Bo$ for the base flow ($Bo_b=\{0.01,0.65\}$) and varying it for perturbations ($Bo_p$);
    (b) the maximal growth rate obtained by varying $Bo$ for the base flow ($Bo_b$) and isolating it for the unstable perturbations ($Bo_p=\{0.01,0.65\})$;
    the varying base interfaces are shown;
    In both panels, thin dashed lines show the tube wall and thin solid lines show the base interfaces, the thick solid lines represent a stable regime, and pentagrams mark the stabilisation thresholds, and the green colour presents the analytical prediction by~\citet{Duclaux2006}.
 }
	\label{fig:internaldrainage_Bop_Bod}
\end{figure}

\indent To shed light on the reasons for the present improvement in the prediction of the stability threshold, we recall from \S\ref{subsec:internaldrainage-baseflow} that the present study exhibits for a high $Bo$ a remarkable deformation of the rising bubble (also shown in figure~\ref{fig:internaldrainage_phasediagram}(a)).
The interface deformation is disregarded in the analysis of~\citet{Duclaux2006}. Yet, their analysis predicts the suppression of the instability. 
To elucidate the stabilising role of the Bond number, we take benefit of the linear instability formalism to decouple artificially the effect of $Bo$ into (i)~its capacity {to} deform the base state as {a} result of drainage and capillary forces, and (ii)~its explicit role in the perturbation equations (\ref{eq:internaldrainage-eigenvalue-problem}) and (\ref{eq:internaldrainage_eigenvalue_operators_global}).
{As detailed below, we will therefore distinguish the "perturbation Bond number " $Bo_p$ and the "base Bond number " $Bo_b$.}
\begin{enumerate}
    \item First, we isolate the effect of the {perturbation} Bond $Bo_p$  on the stability properties by fixing the base state according to two extreme values of {the base Bond number} $Bo_b$. {In other words, we} artificially prevent the Bond number from deforming the base state but let it act {on} the perturbations.
    We consider the two limits of low ($Bo_b=0.01$) and high ($Bo_b=0.65$) base Bond numbers for $\delta/ \beta =0.3$.
    The former is unstable and the latter stable according to {our} linear analysis (figure~\ref{fig:internaldrainage_phasediagram}), and the stabilisation threshold is $Bo \approx 0.575$.
    Fixing the bubble interface to that {obtained} at $t_f$, the effect of $Bo_p$
    is investigated by varying it
    over a wide range.
    Remember that the base flow at $t_f$ is quasi-stagnant. The variation of the maximal growth rate, $\sigma_r^{max}$, is presented in figure~\ref{fig:internaldrainage_Bop_Bod}(a) along with its corresponding prediction from the dispersion relation of~\citet{Duclaux2006} (green) and the "matching" value {obtained when}  $Bo_b=Bo_p$ (black squares).
    The results {show} that the growth rate for the bubble with a circular cross-section ($Bo_b=0.01$, blue circles) diminishes {with} $Bo_p$.
    However, the interface does not stabilise (the growth rate does not cross zero)   even for a  rather large perturbation Bond number, $Bo_p=10^2$.
    In contrast, the deformed interface of $Bo_b=0.65$ (red crosses) stabilises as $Bo_p$ exceeds a threshold (red pentagram) inferior to that of the "matching" 
    case $Bo_b=Bo_p$ (black pentagram), and far 
    {larger than} that of~\citet{Duclaux2006}~(green pentagram).
    Moreover, for small $Bo_b $ where the bubble interface is almost circular, the isolated growth rate (blue circle) is close to the 
    "matching" value (black square).
    We observe that the present linear analysis does not agree with the growth rate {predicted} by~\citet{Duclaux2006} even in the low $Bo$ limit,
    where the interface is circular. 
    It is seen that the base bubble deformation state plays a significant role: a more {deformed} bubble {(larger $Bo_b$)} is less unstable so that a lower {$Bo_p$} number 
    is needed to eventually stabilise the flow than {when} $Bo_p=Bo_b$. 
    {Remember that,} in contrast, a quasi-cylindrical bubble {(smaller $Bo_b$)} is found more unstable than the "matching" case $Bo_p=Bo_b$,
    {and} is actually never stabilised. This demonstrates the essential role of the perturbation Bond number $Bo_p$. 
    \item {Next, we} artificially fix the perturbation Bond number $Bo_p$ to the same extreme values ($Bo_p=0.01$ and $Bo_p=0.65$) and let  the base Bond number $Bo_b$ influence the base state. Not surprisingly, a larger perturbation Bond number ($Bo_p=0.65$, red crosses) is {more} stabilising {than} the "matching" case (black squares), while a smaller perturbation Bond number ($Bo_p=0.01$,  blue circles) is {more} destabilising. The associated {critical} $Bo_b$ numbers (blue and red pentagrams) {bracket} the "matching"  critical Bond number (black pentagram). This demonstrates the role of 
    {the Bond number on the base state and, in turn, on the stability properties.}
\end{enumerate}
\indent To conclude, decoupling the effects of $Bo$ on the base flow and perturbations  suggests that {its influence on the} base state (as revealed by the bubble deformation at $t_f$) is essential to stabilise the flow. 
{Without base state deformation, the instability would persist at large Bond numbers.}

{This remark contrasts with what~\citet{Duclaux2006} concluded, namely that the whole flow would become stable as soon as any region  would become stable (in this case the north pole).}
The mismatch between their analytical threshold and experiments (as well as our analysis) could therefore be possibly explained by the circular bubble cross-section assumption in their derivation of the linear stability characteristics.
Another {source of} discrepancy {may be related to} their dispersion relation, {where} the growth rate is non-{uniform} in $\theta$, whereas {in principle} the normal mode assumption forbids any dependency of the growth rate on the spatial coordinates.
\subsubsection{Maximal unstable wavenumber} \label{subsec:internaldrainage_results_kmax} 
\begin{figure}
	\centerline{\includegraphics[width=1\textwidth]{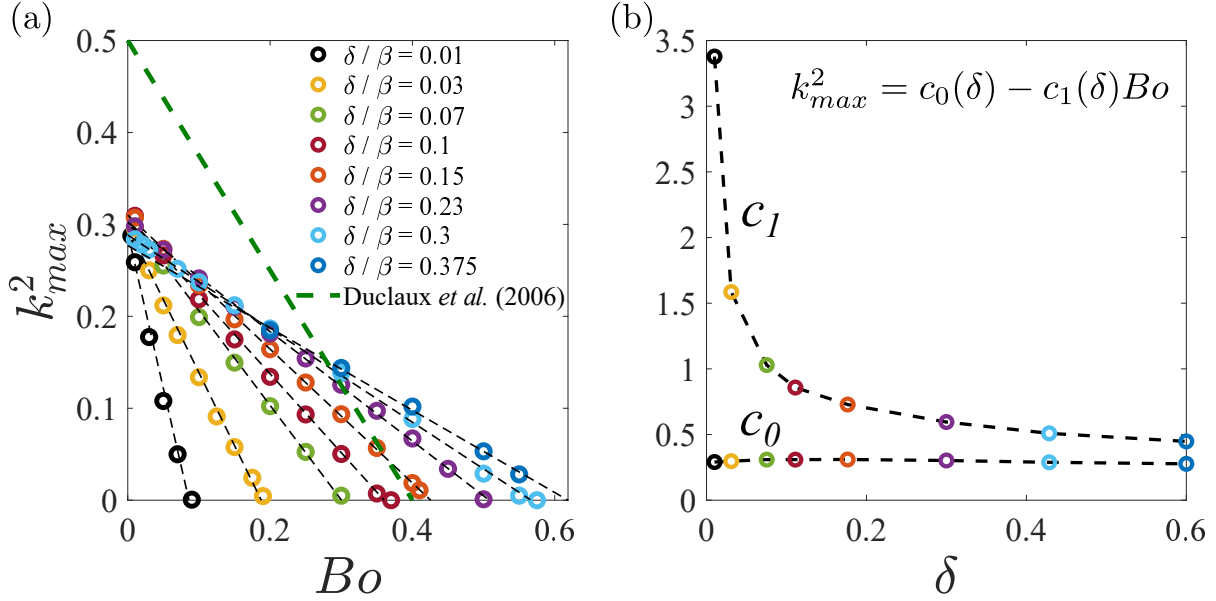}} 
	\caption{Maximal wavenumber ($k_{max}$) obtained from the linear stability analysis: (a) $k_{max}^2$ vs $Bo$ for different initial film thickness values; the black dashed lines show the linear fit to the present numerical data, $k_{max}^2 = c_0(\delta) -c_1(\delta) Bo$, and the green dashed line shows the prediction by~\citet{Duclaux2006}; (b) the prefactors obtained from the fit shown in panel (a).	
 }
	\label{fig:internaldrainage_kmaxdiagram}
\end{figure}

\indent The maximal wavenumber, associated with the maximal growth rate of the unstable mode 1 is shown in figure~\ref{fig:internaldrainage_kmaxdiagram} for a wide range of parameters.
Increasing $Bo$ decreases the optimal wavelength and $k_{max}\rightarrow 0$ on the stability boundary.
While~\citet{Duclaux2006} propose the fit $k_{max}^2 = 0.5 -1.25 Bo$, the present linear analysis evidences a similar trend but with different fitting parameters that depend on the initial film thickness.
These fitting parameters are presented in figure~\ref{fig:internaldrainage_kmaxdiagram}(b). 
Note that considering the universal relation of~\citet{Duclaux2006}, one expects a unique stabilisation threshold of $Bo=0.4$ for all film thickness values in contrast to their phase diagram. 

\subsubsection{Validity of the frozen frame assumption} \label{subsec:internaldrainage_results_frozendic} 
\begin{figure}
	\centerline{\includegraphics[width=1\textwidth]{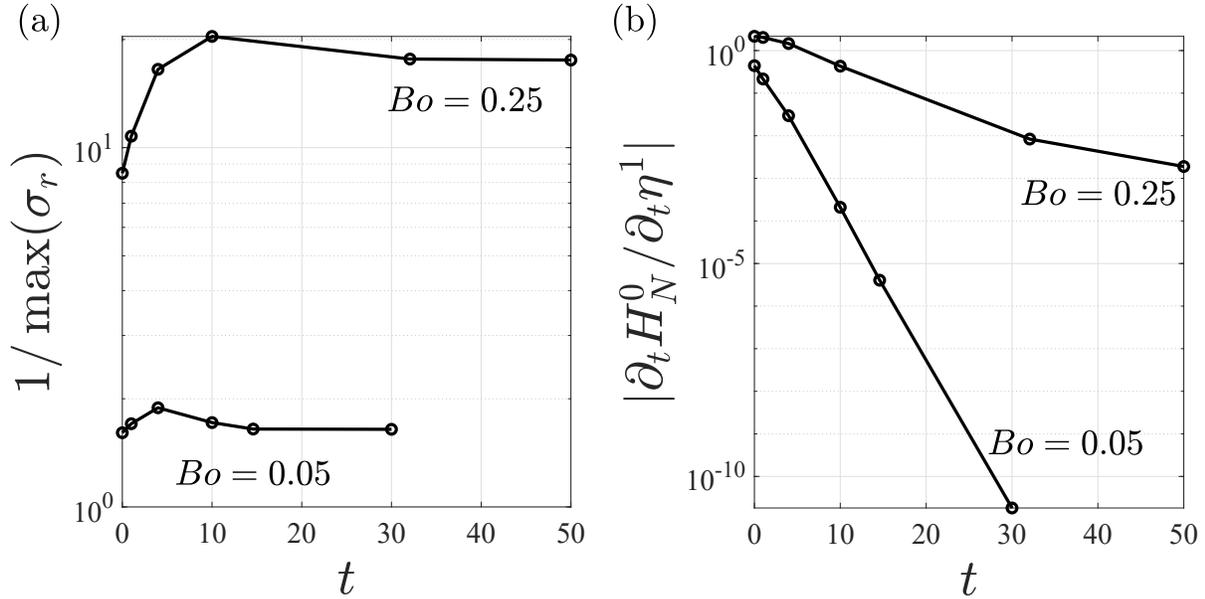}} 
	\caption{Frozen frame analysis validity check for the unstable modes presented in figure~\ref{fig:internaldrainage_Bo_BF}(c):
    (a) Temporal evolution of $1/\max(\sigma_r)$; 
    (b) Comparison of the evolution rate between the base and perturbed flows. 
 }
	\label{fig:internaldrainage_frozenframe}
\end{figure}
We should recall from \S\ref{subsec:internaldrainage-formulation-stability-analysis} that the key assumption behind the frozen frame approach for the linear stability analysis is that the perturbations evolve much faster than the base flow~\citep{Tan1986}.
To assess the validity of this assumption, one can consider the characteristic time scales of the base flow evolution presented in table~\ref{tab:internaldrainage_Bo_effect_BF_fit}.
After the initial bubble drift, the base flow decelerates mainly with two distinct time scales; a fast time scale $T_1 \sim 5$ and a slow time scale $T_2 > 50$.
On the other hand, the relevant time scale for the unstable perturbation's evolution can be computed at different time steps as $1/\max(\sigma_r)$, which is presented in figure~\ref{fig:internaldrainage_frozenframe}(a). 
To indicate how fast the base flow interface evolves compared to the unstable perturbation, $ | \p_t H_N^0 /  \p_t { \eta^1} | \sim \frac{\delta \left( a_1/T_1 \exp[-t/T_1] + a_2/T_2 \exp[-t/T_2] \right) }{ \max(\sigma_r) \exp\left[ \max(\sigma_r) t \right] }$ is presented in figure~\ref{fig:internaldrainage_frozenframe}(b). The numerator is calculated from the double-exponent fit of the base flow, and the denominator is calculated from the temporal growth of the most unstable mode.
A look at this indicator reveals that the base flow and perturbations may evolve at similar rates at small times, and even in some cases that the base flow may evolve faster than the perturbation.
Nevertheless, at large times, when the drainage has decayed, the unstable perturbation grows faster than the base flow. Thus, the frozen frame assumption is only valid for large times, and loses its rigor in the early stages.
We recall that the criterion to distinguish between stable and unstable modes in figure~\ref{fig:internaldrainage_phasediagram} is the growth rate of the modes at large times, sufficient for the frozen frame assumption to hold.
However, it is crucial to properly take into account the base flow evolution at early times and to account for possible nonmodal mechanisms.
For this reason, we carry out a transient growth analysis whose results are presented and discussed in the next section.
\subsection{Transient growth analysis}\FloatBarrier \label{subsec:internaldrainage_results_TG}
\begin{figure} 
\centering
  \begin{subfigure}[b]{0.49\linewidth}
\includegraphics[trim={0cm 0cm 0cm 0cm},clip,width=1\linewidth]{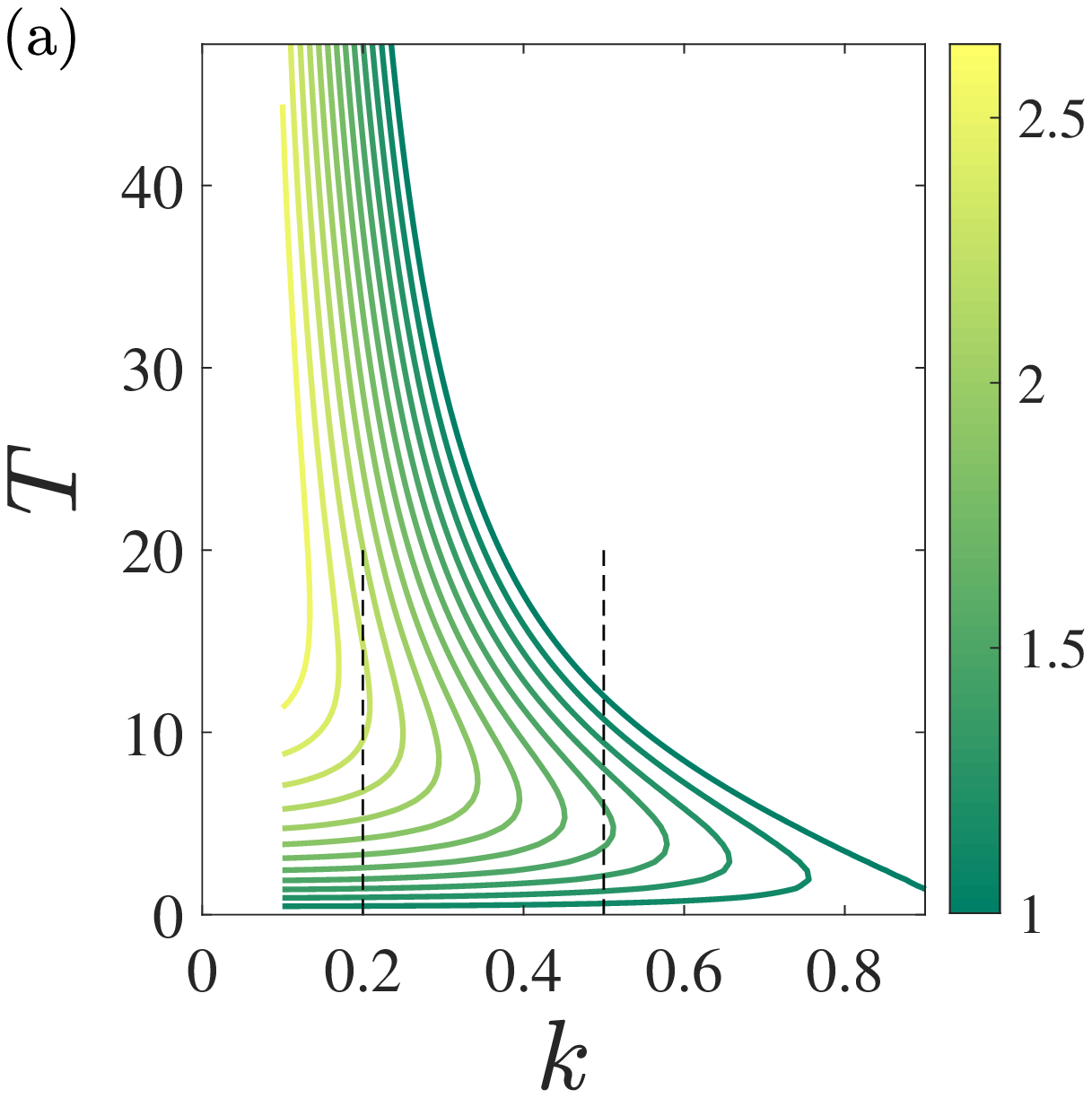}
\caption*{\label{fig:tg1a}}
\end{subfigure}
  \hfill
  \begin{subfigure}[b]{0.49\linewidth}
\includegraphics[trim={0cm 0cm 0cm 0cm},clip,width=1\linewidth]{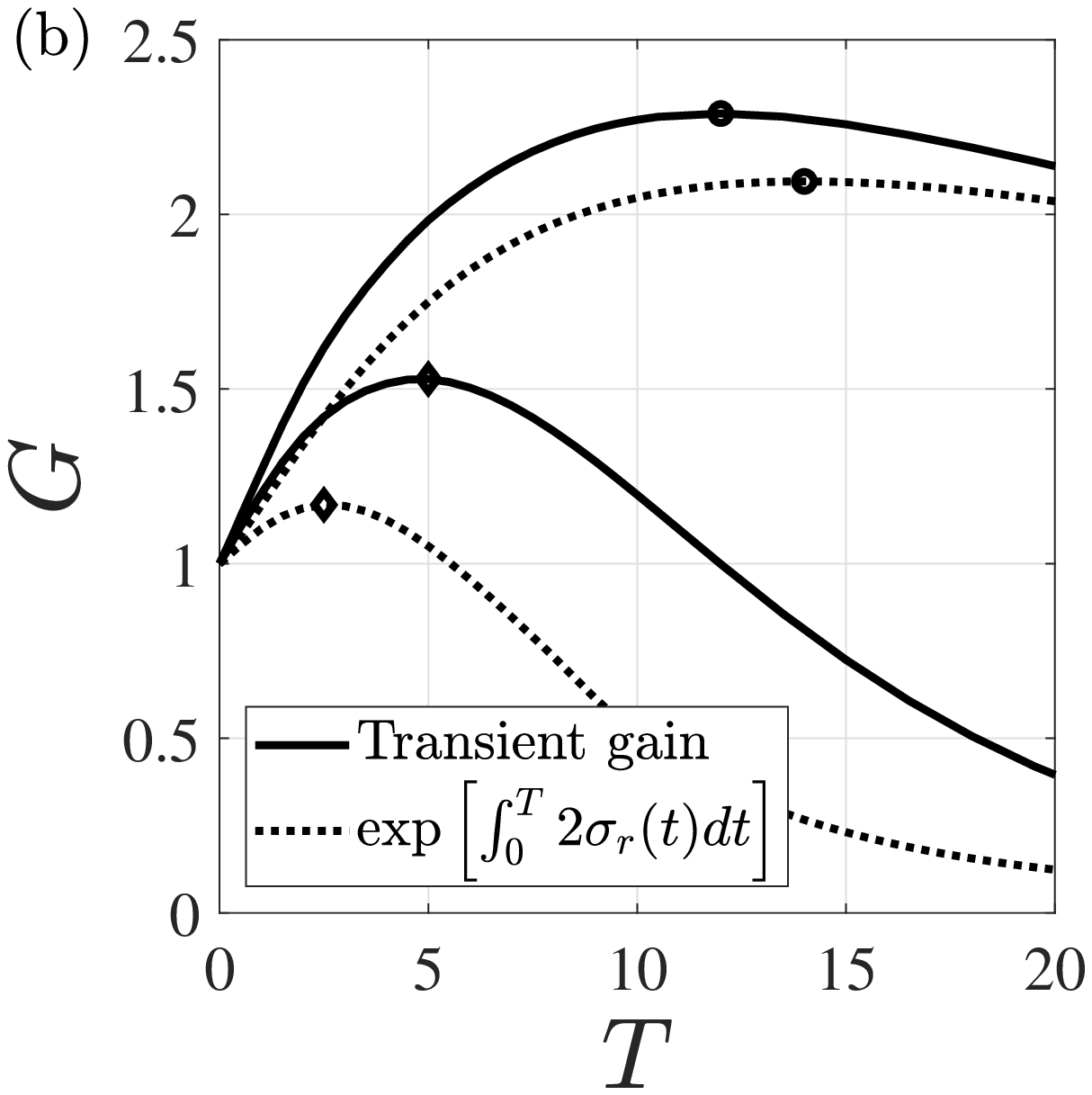}
\caption*{\label{fig:tg1b}}
 \end{subfigure}
\caption{The parameters $\delta/\beta = 0.3$ and $Bo = 0.65$ have been selected. \text{(a)} Isocontours of the transient gain $G$ defined in (\ref{eq:tgdef}) in the $k-T$ plane, the temporal horizon $T\in[0,t_f]$ (only the region where $G\geq1$ is shown). \text{(b)} Along the thin dashed lines in (a), comparison of the transient gain (full line) with its approximation using the dispersion curve only (dotted line), for $k=0.2$ and $k=0.5$ (maximum gain highlighted by a circle and a diamond, respectively). \label{fig:tg1}} 
\end{figure}
\begin{figure} 
\centering
  \begin{subfigure}[b]{0.48\linewidth}
\includegraphics[trim={0cm 0cm 0cm 0.0cm},clip,width=1\linewidth]{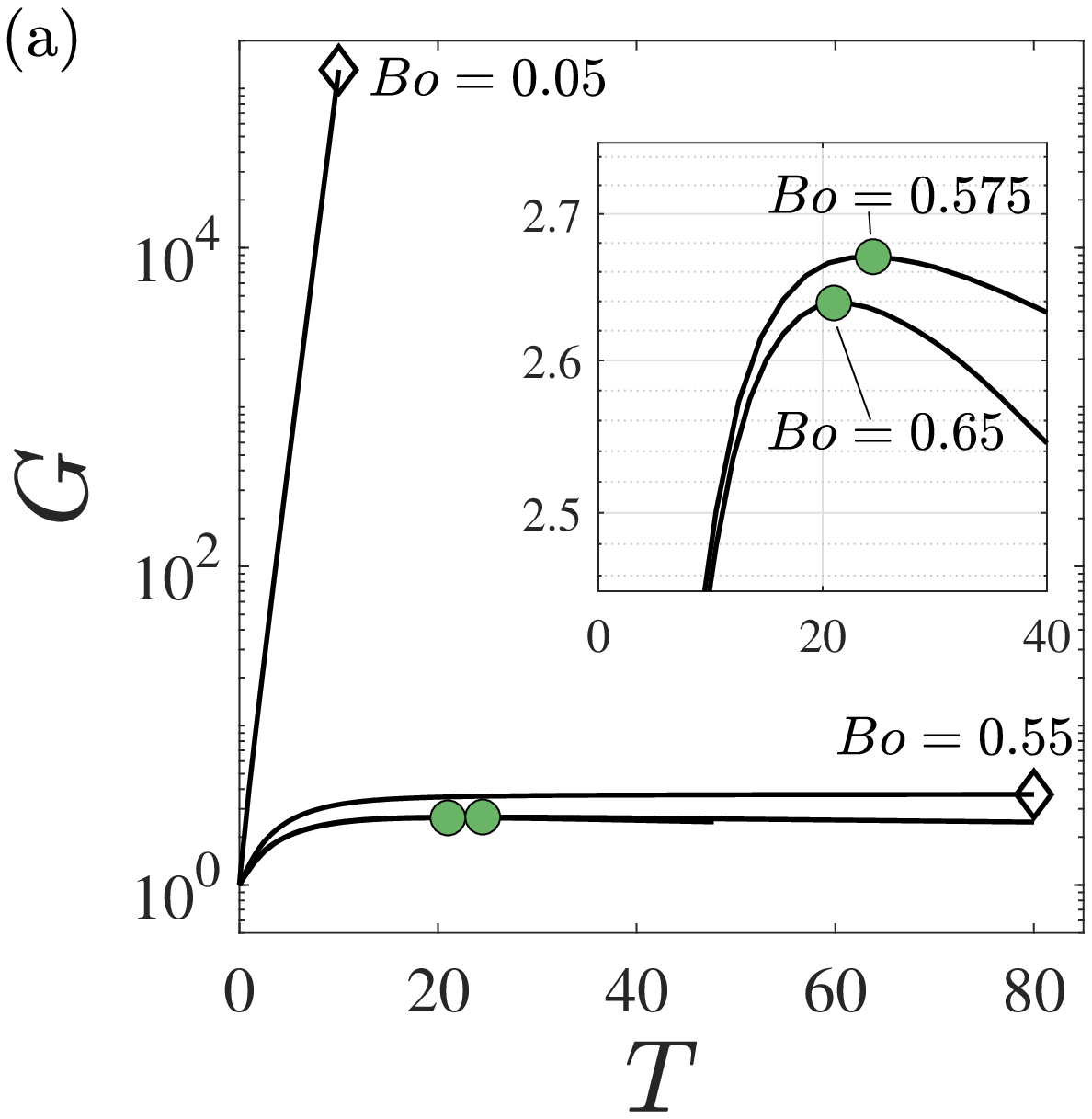}
\caption*{\label{fig:tg2a}}
\end{subfigure}
  \hfill
  \begin{subfigure}[b]{0.49\linewidth}
\includegraphics[trim={0cm 0cm 0cm 0cm},clip,width=1\linewidth]{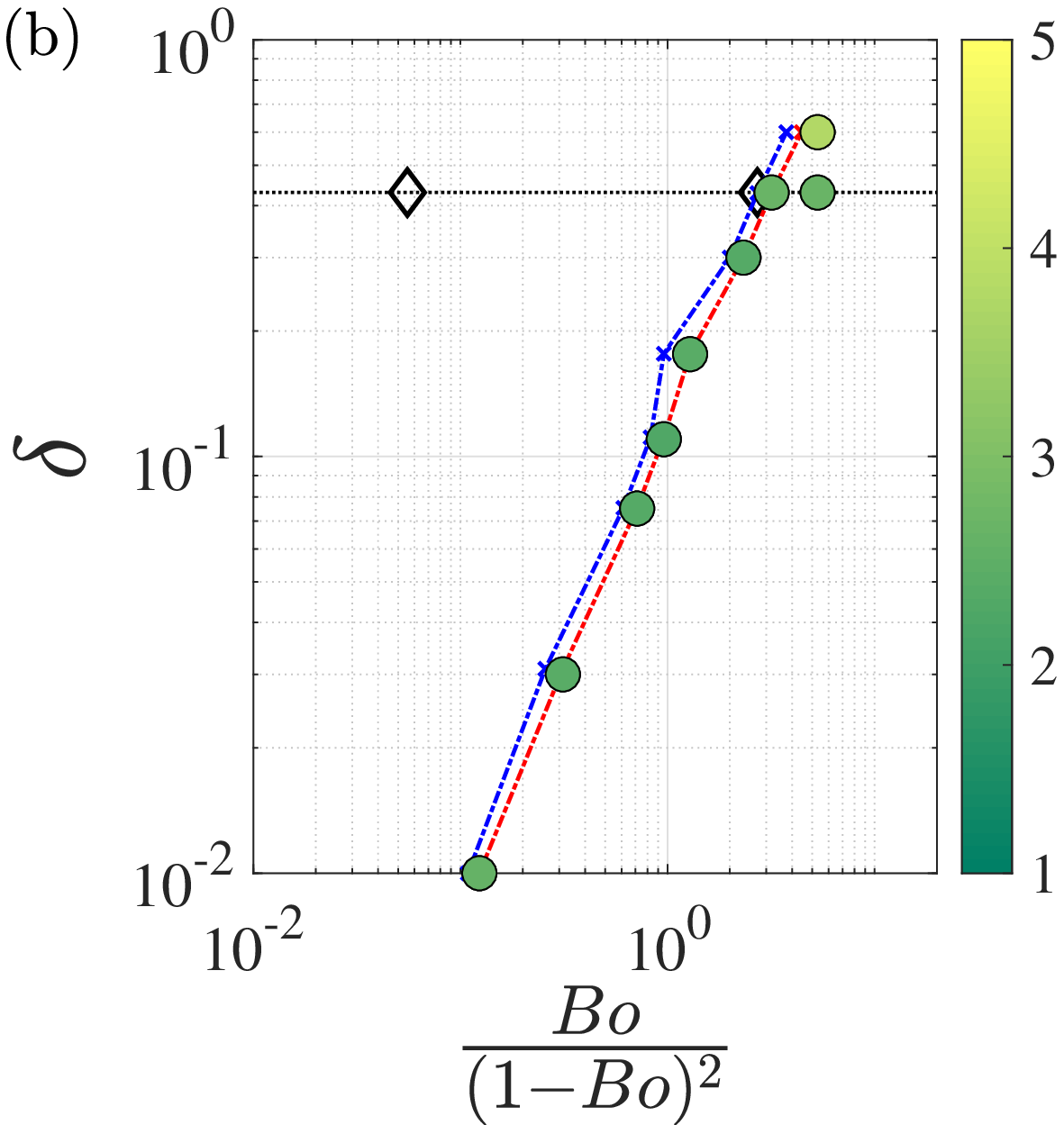}
\caption*{\label{fig:tg2b}}
  \vspace{-0.3cm}
 \end{subfigure}
\caption{{(a)} Transient gains for $\delta=0.43$ (corresponding to $\delta/\beta=0.3$), and $Bo=0.05$ (significantly unstable), $Bo=0.55$ (slightly unstable), $Bo=0.575$ (stable) and $Bo=0.65$ (stable). Bullet (resp. diamond) markers highlight the largest gain for linearly stable (resp. unstable) parameters. Inset: zoom highlighting the largest gains for $Bo=0.575$  and $Bo=0.65$. The wavenumber $k$ leading to the largest growth has been selected for all curves. {(b)} Phase diagram obtained from the transient growth analysis. Colours stand for $\Gm=\max_{k,T}G$ with $k\in[0.1,0.9]$ and $T\in[0,t_f]$. The thin dotted line corresponds to $\delta/\beta=0.3$. The blue (resp. red) dashed-dotted line links the last unstable (resp. first stable) points for a given $\delta$ and by increasing $Bo$.\label{fig:tg2}}
\end{figure}
We present in what follows the results of the transient growth analysis, conducted mainly in the stable part of the parameter space in figure~\ref{fig:internaldrainage_phasediagram}. Indeed, a flow can be linearly (asymptotically) stable but transiently amplify an initial perturbation so significantly that it is destabilised through a subcritical transition. 

The transient gain for $\delta/\beta = 0.3$ and $Bo = 0.65$ is shown in figure~\ref{fig:tg1}(a) as a function of the wavenumber $k$ and the temporal horizon $T$. Longer waves are subject to larger maximum transient gain but require more time before reaching it. The maximum attainable transient gain over all considered $k$ and $T$ is $\Gm=2.64$, reached for the lowest considered $k$. This maximum gain value is of order unity, such that, for the considered parameters at least, a small-amplitude initial perturbation is not expected to be sufficiently amplified by the flow to trigger nonlinearities. 

The extent to which the transient growth is driven by nonmodal mechanisms, and may or may not be estimated directly from dispersion curves is investigated in figure~\ref{fig:tg1}(b), where the transient gains are compared with the integrated growth rate $\exp\left[\int_{0}^{T}2 \sigma_r(k,t)dt\right]$ for $k=0.2$ and $k=0.5$ (thin dashed lines). We recall that the growth rate $\sigma_r(k,t)$ for $\delta/\beta = 0.3$ and $Bo = 0.65$ is shown in figure~\ref{fig:internaldrainage_Bo_BF}(c). For $k=0.5$ in figure~\ref{fig:tg1}(b), the curves are qualitatively similar but present some small quantitative discrepancies: the maximum transient gain reaches a slightly larger value at a larger temporal horizon, $T=5$ against $T=2.5$. Thus, between these two times, the energy of the perturbation can grow although the frozen base flow is linearly stable. For $k=0.2$ the transient gain is also systematically larger, however the two curves present little differences, implying this transient gain to be driven mostly by modal mechanisms. 

In figure~\ref{fig:tg2}(a), the gain for $\delta/\beta=0.3$ and $Bo=0.65$ is compared to those of three additional Bond numbers: $Bo=0.575$, $Bo=0.55$ and $Bo=0.05$, which correspond to a stable, a slightly and a strongly unstable flow, respectively. The wavenumber associated with the largest gain was selected. For $Bo=0.05$, the maximum growth rate remains large (see figure~\ref{fig:internaldrainage_Bo_BF}(c)) and the gain quickly aligns on the fast exponential growth; specifically, the slope of the line marked with the black diamond is two times the maximum growth rate. For $Bo=0.55$, the maximum growth rate rapidly converges towards very small values but does not become negative, thus the gain also grows exponentially for large times. Overall, results shown in figure~\ref{fig:tg2}(a) suggest that large gains can only be attained through instability. 

To assess the generality of this conclusion, we report  $\Gm\doteq\max_{k,T}G$ for $k\in[0.1,0.9]$ and $T\in[0,t_f]$ in figure~\ref{fig:tg2}(b), in the same $\{\delta,Bo\}$ parameter space as in figure~\ref{fig:internaldrainage_phasediagram}. The considered points were voluntarily chosen in the stable regime and the closest to the separatrix (i.e., the first red dot for a given $\delta$ in figure~\ref{fig:internaldrainage_phasediagram}(b)). For all the considered parameters, $\Gm$ does not exceed the small value of $\Gm=3.8$. In this sense, the linear stability analysis seems to provide a sufficiently complete description of this flow, since linear stability seems to imply small transient gains; conversely, linear instability seems to be a necessary and sufficient condition for large gains. Moreover, as we saw for $\delta/\beta=0.3$ and $Bo=0.65$, and we have checked these conclusions to be true for all the other points in figure~\ref{fig:tg2}(b), $\Gm$ is reasonably well characterised using solely the information about the growth rate. In conclusion, nonmodal mechanisms are believed to have little influence on the flow, at least for the parameters considered in this paper.



\section{Summary and conclusion} \label{sec:internaldrainage_conclusion}\FloatBarrier
In this work, we have studied the draining flow of a viscous liquid film coating the inner wall of a horizontal tube.
First, the temporal evolution of the axially invariant base flow with an initially stagnant film of uniform thickness was computed numerically.
In the absence of inertia, the base flow exhibits an instantaneous upward drift of the core bubble, followed by an exponential decay of the flow as the interface approaches the tube upper wall. 

Next, the stability of the evolving base flow was investigated by means of a linear stability
analysis under the frozen frame assumption.
It was verified that this assumption holds at large times.
One unstable mode was observed which features the characteristics of the Rayleigh-Plateau mode: axial interface undulations, horizontal symmetry, and vertical asymmetry evidencing stronger modulation at the bottom.
These features match the interface shape observed in the experiments of ~\citet{Duclaux2006}. The maximal growth rate of this mode decreases 
as the bubble approaches the tube upper wall.

A parametric study was then conducted in the space of dimensionless parameters $\{Bo, Oh, \beta \}$. 
This study suggests that increasing $Bo$, equivalent to weakening  surface tension 
{in comparison with} gravity, results in a horizontally wider bubble interface 
{that rises more slowly} 
than quasi-cylindrical bubbles observed at small Bond numbers. 
The bubble slowdown coincides with the 
displacement of the {location of} minimum film thickness from the north pole to the sides.
{With increasing $Bo$, the deformed interface becomes less unstable, and the Rayleigh-Plateau instability is suppressed above a critical $Bo$ value.}
We {also} demonstrated that  inertial forces, achieved at finite $Oh$, affect the flow mainly at the onset of the drainage such that the bubble drifts upward {more} smoothly, and the drainage is delayed.
However, inertia does not alter the shape of the interface, nor its linear stability regime, and has a minor influence on {the} maximal growth rate at large times.

{Finally,} 
{a stability} diagram was sketched by investigating the linear stability of the deformed interface in the limit of large times for various $\{ Bo,\beta \}$,
{confirming the stabilising effect of larger Bond numbers and thinner films.}
{By relaxing assumptions of past studies (e.g. circular base interface), the present linear study showed an interesting improvement on existing theoretical results for the transition between stable and unstable interfaces. A slight discrepancy with the experimental data of~\citet{Duclaux2006} remains, resulting in a few stable experimental conditions predicted to be unstable by our linear stability analysis.}
Comparing the experimental protocol of~\citet{Duclaux2006} with the assumptions of the present study suggests some possibilities 
for the observed slight mismatch. 
Firstly, 
{the most unstable} wavenumber diminishes {with $Bo$}, which results in the promotion of very long wavelengths. 
Capturing
{wavelengths longer than the finite length of the experimental apparatus is not possible, which may result in the stabilization of perturbations}
that would be unstable in an infinitely long tube.
Additionally, in the experiments, the gaseous core is pushed into a wet tube by a syringe pump;
{any residual} axial velocity, 
left-right asymmetry, or axial {variations in} the deposited film thickness~\citep{Balestra20183} may all have some influence on the stability.
Finally, we note that linear analysis cannot capture the influence of non-linearities on the final pattern. Non-linearities may quickly saturate the growth of the linear modes to a very small amplitude~\citep{Halpern2003} which may make it difficult to {observe} the instability {experimentally}.

Lastly, a transient growth analysis was conducted so as to relax the frozen base flow assumption. It 
{conclusively demonstrated} the small importance of nonmodal mechanisms, both because the transient gains were systematically of order unity in the linearly stable region of the parameter space, and because they were rather well predicted from the growth rate of the leading eigenvalue. This justifies the relevance of the linear stability analysis for this flow when compared with experimental data. 

{The present study, based on the full Navier-Stokes equations and capable of handling complex interface geometries, paves the way for a wealth of future investigations.
For instance, it would be of interest to investigate the effect of inclining the tube: as gravity becomes non-orthogonal to the tube axis, one can expect a competition between transverse drainage and longitudinal advection, possibly resulting in a transition from absolute to convective instability. 
Another exciting perspective is that of a film coating the outside of an inclined tube: in addition to the aforementioned competition, this configuration offers the possibility of rich nonlinear dynamics such as pinch-off.
Finally we recall that the present study has focused on the regime of small and intermediate Bond numbers, where the Rayleigh-Taylor instability is suppressed~\citep{Trinh2014,Balestra2016}. A natural extension of our work should explore larger Bond numbers. We conjecture the existence of an interval of stable Bond numbers, with smaller $Bo$ unstable to the Rayleigh-Plateau instability (like in the present study) and larger $Bo$ unstable to the Rayleigh-Taylor instability~\citep{Trinh2014,Balestra2016,Balestra20182}.}






\newpage
\appendix
\section{Derivation of the interface boundary conditions}\label{app:internaldrainage_interfacebc}\FloatBarrier
In this section, the derivation of the interface boundary conditions for the perturbed flow is elaborated. These conditions should be imposed on the perturbed interface, i.e. on $r=\Rb^0 +\epsilon \eta^1$, while $\eta^1$ is already a part of the problem unknowns. By using the Taylor expansion, that is, projecting radially on the base interface, i.e. on $r=\Rb^0(\theta,t)$, any flow quantity at the perturbed interface can be readily approximated. This projection is referred to as flattening and for an arbitrary function $f(r,\theta,z,t)$ can be expressed as
\begin{equation} \label{eq:internaldrainage-flattening}
	f|_{(r=\Rb^0+ \epsilon \eta^1 ,\theta,z,t )} = f|_{(r=\Rb^0,\theta,z,t )} + \epsilon \eta^1 \ \p_r f|_{(r=\Rb^0,\theta,z,t )} + \mathcal{O}(\epsilon^2) .
\end{equation}

By substituting the decomposed state vector of~(\ref{eq:internaldrainage-Solution-decomposition}), into the interface conditions~(\ref{eq:internaldrainage-kinematic-BC})-(\ref{eq:internaldrainage-dynamic-BC}), then using the ansatz of~(\ref{eq:internaldrainage_eigenmode_ansatz}), and applying the aforementioned flattening, we can formulate these conditions as a set of equivalent constraints on the boundary of the base interface. The linearised form of the kinematic condition~(\ref{eq:internaldrainage-kinematic-BC}) writes

\begin{equation} \label{eq:internaldrainage-kinematic-BC-lin1}
	\p_t \left( \Rb^0+ \epsilon \eta^1 \right) + \left( {\bf u}^0 + \epsilon {\bf u}^1 \right)\cdot \nabla \left( \Rb^0+ \epsilon \eta^1 \right) = \left( {\bf u}^0 + \epsilon {\bf u}^1   \right) {\bf \cdot e}_r \quad \text{ at } \ r=\Rb^0 + \epsilon \eta^1,
\end{equation}
where the gradient vector in the Cylindrical coordinate can be expressed as $\nabla = \left( \p_r, 1/r \p_\theta, \p_z \right)^T$. 
Applying~(\ref{eq:internaldrainage-flattening}) to~(\ref{eq:internaldrainage-kinematic-BC-lin1}) and using the ansatz of~(\ref{eq:internaldrainage_eigenmode_ansatz}) readily results in~(\ref{eq:internaldrainage_Interface_BC_linearized_kin}).

The linearised dynamic condition~(\ref{eq:internaldrainage-dynamic-BC}) writes 
\begin{equation} \label{eq:internaldrainage-dynamic-BC-lin1}
	\left( \stau^0 + \epsilon \stau^1 \right) \ \left( {\bf n}^0 + \epsilon {\bf n}^1 \right)  = \left( \kappa^0 + \epsilon \kappa^1 \right) \left( {\bf n}^0 + \epsilon {\bf n}^1 \right)  \quad \text{ at } \ r=\Rb^0 + \epsilon \eta^1,
\end{equation}
Applying~(\ref{eq:internaldrainage-flattening}) to~(\ref{eq:internaldrainage-dynamic-BC-lin1}) and using the ansatz of~(\ref{eq:internaldrainage_eigenmode_ansatz}) readily results in~(\ref{eq:internaldrainage_Interface_BC_linearized_dyn}). In order to express interface conditions in the Cartesian coordinates, the terms which are expressed in the Cylindrical coordinates should be transformed by employing the Jacobian transformations as 
\begin{align}
	{\bf e}_r &= \cos{\theta} \ {\bf e}_x + \sin{\theta} \ {\bf e}_y, & {\bf e}_{\theta} &= - \sin{\theta} \ {\bf e}_x + \cos{\theta} \ {\bf e}_y, \nonumber\\
	\p_r &= \cos{\theta} \ {\p}_x + \sin{\theta} \ {\p}_y, &\p_{\theta} &=  \frac{{\bf t}^0 \cdot \nabla_s}{{\bf t}^0 \cdot \nabla_s \theta },
\end{align}
where ${\bf t}^0$ denotes the unit tangent vector, and $\nabla_s = \nabla - {\bf n}^0 \left( {\bf n}^0 \cdot \nabla \right) $ is the tangential derivative on the base interface. Both conditions~(\ref{eq:internaldrainage_Interface_BC_linearized_kin}) and (\ref{eq:internaldrainage_Interface_BC_linearized_dyn}) include the normal vector and the curvature of the perturbed interface whose formulation is given in appendix~\ref{app:internaldrainage_interfacecharacterization}. For further details concerning the numerical implementation of the boundary conditions, see appendix~\ref{app:internaldrainage_implementation_LSA}.

\section{Variational formulation of the linear stability analysis and implementation of boundary conditions}\label{app:internaldrainage_implementation_LSA} 
\begin{figure}
		\centerline{\includegraphics[width=1\textwidth]{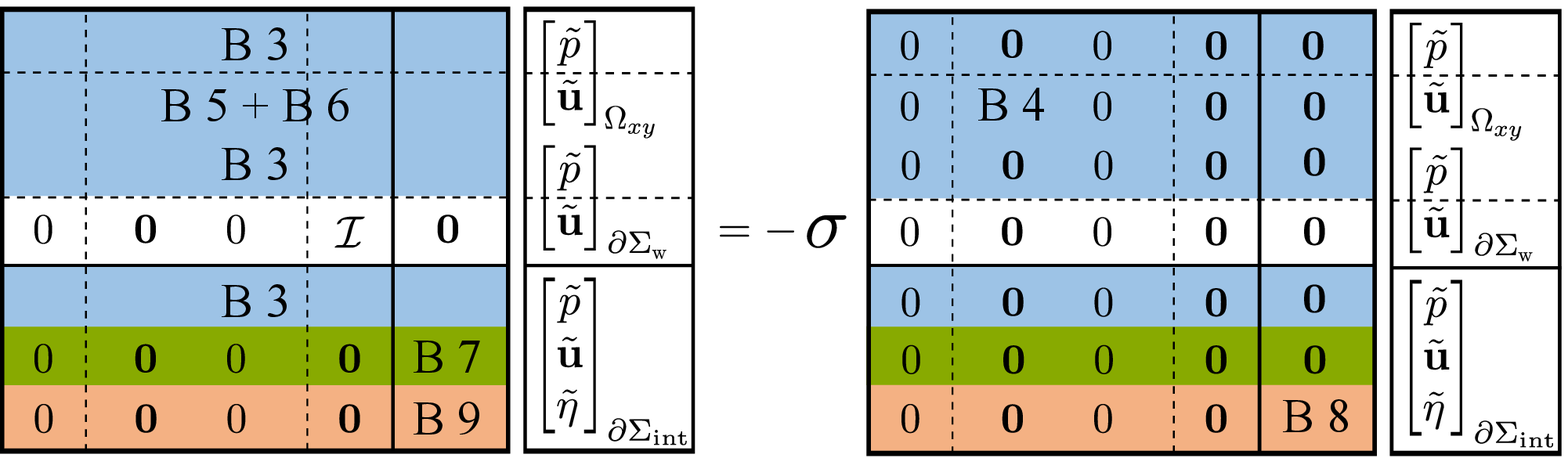}}
	\caption{Matrix representation of the variational system~(\ref{eq:internaldrainage_weak3_LSA_1})-(\ref{eq:internaldrainage_weak3_LSA_5}), solved in COMSOL Multiphysics$^\text{TM}$; blue represents the implementation of~(\ref{eq:internaldrainage-linearized-Incompressibility})-(\ref{eq:internaldrainage-linearized-momentum}); white represents the implementation of the no-slip boundary condition on the solid wall; green represents the implementation of the dynamic boundary condition~(\ref{eq:internaldrainage_Interface_BC_linearized_dyn}); a represents the implementation of the kinematic condition~(\ref{eq:internaldrainage_Interface_BC_linearized_kin}).}
	\label{fig:internaldrainage_numerics}
\end{figure}

Implementation of the numerical scheme and development of the {\it variational formulation} associated with the governing equations presented in section~\ref{sec:internaldrainage-Gov-eq} are elaborated in this appendix, recalling that the numerical domain is shown in figure~\ref{fig:internaldrainage_numerical_domain_drainage_cartesian}. To develop the variational form of~(\ref{eq:internaldrainage-eigenvalue-problem}), firstly the normal mode of~(\ref{eq:internaldrainage_eigenmode_ansatz}) is applied to the system of equations~(\ref{eq:internaldrainage-linearized-Incompressibility})-(\ref{eq:internaldrainage_Interface_BC_linearized_kin}). Then it is internally multiplied by the vector of the test functions $\psi=[\psi_{p},\psi_{{\bf u}},\psi_{\eta}]$, where $\psi_{{\bf u}}=[ \psi_{u_x},\psi_{u_y},\psi_{u_z}]$. The resulting scalar product is integrated on $\Omega_{xy}$, which in the linear order gives
\begin{align} \label{eq:internaldrainage_weak1_LSA}
	& \biggl\{ \iint_{\Omega_{xy}} \psi_{p}^\star \left( \tilde{\nabla} \cdot \tilde{{\bf u}} \right) \ \mathrm{d} A_{\Omega_{xy}}  \nonumber \\
	+& \iint_{\Omega_{xy}} \psi_{{\bf u}}^\star \cdot  \left( \left( \frac{Bo}{Oh} \right)^2 \delta^4 \left( \sigma \tilde{\bf u} + {\bf u}^0 \cdot \tilde{\nabla} \tilde{\bf u} + \tilde{\bf u} \cdot \nabla {\bf u}^0 \right) - \tilde{\nabla} \cdot \underline{\underline{\tilde{\tau } }} \right) \ \mathrm{d} A_{\Omega_{xy}}  \nonumber \\
	+&  \int_{ \p \Sigma_\mathrm{int} } \psi_{\eta}^\star \left( 	\sigma \tilde{\eta} + \left(  -\p_r u^0_r + \frac{\p_r u^0_\theta \ \p_\theta \Rb^0}{\Rb^0} - \frac{u^0_\theta \ \p_\theta \Rb^0 }{\left(\Rb^0\right)^2} \right) \tilde{\eta} + \frac{u^0_\theta  }{\Rb^0} \p_\theta \tilde{\eta}  + \ \frac{\p_\theta \Rb^0}{\Rb^0}  \tilde{u}_{\theta} - {\tilde{u}_r } \right)  \ \mathrm{d}s \biggr\}  \nonumber \\
	+& \text{c.c.}= 0.
\end{align}
It should be noted that in a complex system, the applied scalar product is Hermitian, defined as $\left\langle {\bf a , b} \right\rangle = {\bf a^\star \cdot b}$ where the superscript $\star$ denotes the complex conjugate. In the last line of this system of equations, kinematic condition~(\ref{eq:internaldrainage_Interface_BC_linearized_kin}) is used to define $ \tilde{\eta}$ only on \(\p\Sigma_\mathrm{int}\). After integrating by part, $ \psi_{{\bf u}}^\star \cdot \left(  \tilde{\nabla} \cdot  \underline{\underline{\tilde{\tau } }} \right)  = \tilde{\nabla} \cdot \left(   \underline{\underline{\tilde{\tau } }} \psi_{{\bf u}}^\star \right) - tr\left(   \underline{\underline{\tilde{\tau } }}^T \left( \tilde{\nabla} \psi_{{\bf u}} \right)^\star \right) $ , and then applying the Gauss's theorem,  $\iint_{\Omega_{xy}} \tilde{\nabla} \cdot \left(   \underline{\underline{\tilde{\tau } }} \psi_{{\bf u}}^\star \right) \ \mathrm{d} A_{\Omega_{xy}} = \int_{\p \Omega_{xy}}  \left(   \underline{\underline{\tilde{\tau } }} \psi_{{\bf u}}^\star \right) \cdot {\bf n}^0 \ \mathrm{d}s$,~(\ref{eq:internaldrainage_weak1_LSA}) implies 
\begin{align} \label{eq:internaldrainage_weak2_LSA}
	& \biggl\{ \iint_{\Omega_{xy}} \psi_{p}^\star \left( \tilde{\nabla} \cdot \tilde{{\bf u}} \right) \ \mathrm{d} A_{\Omega_{xy}} \nonumber \\
	+& \iint_{\Omega_{xy}} \psi_{{\bf u}}^\star \cdot  \left( \left( \frac{Bo}{Oh} \right)^2 \delta^4 \left( \sigma \tilde{\bf u} + {\bf u}^0 \cdot \tilde{\nabla} \tilde{\bf u} + \tilde{\bf u} \cdot \nabla {\bf u}^0 \right)  \right) \ \mathrm{d} A_{\Omega_{xy}} \nonumber \\ 
	+& \iint_{\Omega_{xy}} tr\left(   \underline{\underline{\tilde{\tau } }}^T  \left( \tilde{\nabla} \psi_{{\bf u}} \right)^\star \right) \ \mathrm{d} A_{\Omega_{xy}}  \nonumber \\
	+& \int_{\p \Omega_{xy}}  - \left(   \underline{\underline{\tilde{\tau } }} \psi_{{\bf u}}^\star \right) \cdot {\bf n}^0 \ \mathrm{d}s \nonumber \\
	+&  \int_{ \p \Sigma_\mathrm{int} } \psi_{\eta}^\star \left( 	\sigma \tilde{\eta} + \left(  -\p_r u^0_r + \frac{\p_r u^0_\theta \ \p_\theta \Rb^0}{\Rb^0} - \frac{u^0_\theta \ \p_\theta \Rb^0 }{\left(\Rb^0\right)^2} \right) \tilde{\eta} + \frac{u^0_\theta  }{\Rb^0} \p_\theta \tilde{\eta}  + \ \frac{\p_\theta \Rb^0}{\Rb^0}  \tilde{u}_{\theta} - {\tilde{u}_r } \right)  \ \mathrm{d}s \biggr\}  \nonumber \\
	+& \text{c.c.} = 0.
\end{align}
$\underline{\underline{\tilde{\tau } }}$ is symmetric, thus $\left(   \underline{\underline{\tilde{\tau } }} \psi_{{\bf u}}^\star \right) \cdot {\bf n}^0 = \left(   \underline{\underline{\tilde{\tau } }} {\bf n}^0 \right) \cdot \psi_{{\bf u}}^\star$. Using the dynamic condition~(\ref{eq:internaldrainage_Interface_BC_linearized_dyn}) and the fact that $\psi_{{\bf u}}|_{ \p \Sigma_\mathrm{f} } = 0$ (because of the no-slip condition on the solid wall), the variational form of~(\ref{eq:internaldrainage-eigenvalue-problem}) implies
\begin{align} 
	& \biggl\{  \iint_{\Omega_{xy}} \psi_{p}^\star \left( \tilde{\nabla} \cdot \tilde{{\bf u}} \right) \ \mathrm{d} A_{\Omega_{xy}}  \label{eq:internaldrainage_weak3_LSA_1}\\
	+& \iint_{\Omega_{xy}} \psi_{{\bf u}}^\star \cdot  \left( \left( \frac{Bo}{Oh} \right)^2 \delta^4 \sigma \tilde{\bf u}   \right) \ \mathrm{d} A_{\Omega_{xy}}  \\
	+& \iint_{\Omega_{xy}} \psi_{{\bf u}}^\star \cdot  \left( \left( \frac{Bo}{Oh} \right)^2 \delta^4 \left(  {\bf u}^0 \cdot \tilde{\nabla} \tilde{\bf u} + \tilde{\bf u} \cdot \nabla {\bf u}^0 \right)  \right) \ \mathrm{d} A_{\Omega_{xy}} \\ 
	+& \iint_{\Omega_{xy}} tr\left(   \underline{\underline{\tilde{\tau } }}^T  \left( \tilde{\nabla} \psi_{{\bf u}} \right)^\star \right) \ \mathrm{d} A_{\Omega_{xy}}   \label{eq:internaldrainage_weak3_LSA_3}\\
	+& \int_{ \p \Sigma_\mathrm{int} }  \left(   \stau^0 \  \tilde{\bf {n}} + \tilde{\eta}  \, \p_r \stau^0 \  {\bf{n}}^0 - \left( \kappa_0 \tilde{\bf n} + \tilde{\kappa} {\bf{n}}^0 \right) \right) \cdot \psi_{{\bf u}}^\star \ \mathrm{d}s  \label{eq:internaldrainage_weak3_LSA_4}\\
	+&  \int_{ \p \Sigma_\mathrm{int} } \psi_{\eta}^\star \left( \sigma  \tilde{\eta}  \right) \ \mathrm{d}s 	\\
	+&  \int_{ \p \Sigma_\mathrm{int} } \psi_{\eta}^\star \left( 	\left(  -\p_r u^0_r + \frac{\p_r u^0_\theta \ \p_\theta \Rb^0}{\Rb^0} - \frac{u^0_\theta \ \p_\theta \Rb^0 }{\left(\Rb^0\right)^2} \right) \tilde{\eta} + \frac{u^0_\theta  }{\Rb^0} \p_\theta \tilde{\eta}  + \ \frac{\p_\theta \Rb^0}{\Rb^0}  \tilde{u}_{\theta} - {\tilde{u}_r } \right)  \ \mathrm{d}s \biggr\}  \\
	+& \text{c.c.} = 0.	\label{eq:internaldrainage_weak3_LSA_5}
\end{align}
This variational equation can be readily implemented and solved in COMSOL Multiphysics$^\text{TM}$. It is sufficient to solve the first part (in $\{ \}$) and the c.c. is known consequently. The matrix representation of~(\ref{eq:internaldrainage_weak3_LSA_1})-(\ref{eq:internaldrainage_weak3_LSA_5}) is shown in figure~\ref{fig:internaldrainage_numerics}. 

\section{Characterization of an arbitrary interface}\label{app:internaldrainage_interfacecharacterization}
In this section, we present the geometrical characterization of an arbitrary interface parameterised in Cylindrical coordinates. The aim is to develop the characteristics of a three-dimensional interface, as well as the linear perturbations applied to this interface. The key properties of interest are the normal vector and the local curvature. 

\subsection{Normal vector}\label{app:internaldrainage_interfacecharacterization_normal}
We recall that the unit normal vector of a linearly perturbed interface can be decomposed as ${\bf n} = {\bf n}^0 + \epsilon {\bf n}^1$ which reads

\begin{equation} \label{eq:internaldrainage-normal-base}
	{\textbf{n}}^0 = \begin{pmatrix} n^0_r   \\ n^0_\theta   \\  n^0_z \end{pmatrix} = A^{-1/2} \begin{pmatrix} 1   \\ -\frac{1}{\Rb^0} \p_{\theta} {\Rb^0}  \\  \p_{z} {\Rb^0} \end{pmatrix},
\end{equation}

\begin{equation} \label{eq:internaldrainage-normal-linearized}
	{\bf n}^1 = \begin{pmatrix} n^1_r   \\ n^1_\theta   \\  n^1_z \end{pmatrix} =
	\begin{pmatrix} B_{r} \eta^1 + C_{r} \p_\theta \eta^1 + D_{r} \p_{z} \eta^1  \\ B_{\theta} \eta^1 +  C_{\theta} \p_\theta \eta^1 + D_{\theta} \p_{z} \eta^1 \\  B_{z} \eta^1 + C_{z} \p_\theta \eta^1 + D_{z} \p_{z} \eta^1  \end{pmatrix},
\end{equation}
where
\begin{align} 
	\nonumber A &= 1 + \left( \frac{1}{\Rb^0} \p_{\theta} {\Rb^0} \right)^2 + \left( \p_{z} {\Rb^0} \right)^2, &B_r &=  \frac{A^{-3/2}}{\left({\Rb^0}\right)^3} \left( \p_\theta {\Rb^0} \right)^2,  \\
	\nonumber B_\theta &= A^{-3/2}\left( -\frac{1}{\left({\Rb^0}\right)^4} \left( \p_\theta {\Rb^0} \right)^3 + \frac{A}{\left({\Rb^0}\right)^2} \p_\theta {\Rb^0} \right), &B_z &=  -\frac{A^{-3/2}}{\left({\Rb^0}\right)^3} \left( \p_\theta {\Rb^0} \right)^2 \ \p_{z} {\Rb^0},
\end{align}
\begin{align}\label{eq:internaldrainage-normal-lin-coef}
	\nonumber	C_r &=  - \frac{A^{-3/2}}{\left({\Rb^0}\right)^2} \p_\theta {\Rb^0} , & C_\theta &= -\frac{A^{-3/2}}{{\Rb^0}} \left( 1 + \left( \p_z {\Rb^0} \right)^2 \right) , & C_z &=  \frac{A^{-3/2}}{\left({\Rb^0}\right)^2} \p_\theta {\Rb^0} \ \p_z {\Rb^0}, \\  
	D_r &=  -A^{-3/2} \p_{z} {\Rb^0},	 & D_\theta &=   \frac{A^{-3/2}}{{\Rb^0}} \p_{z} {\Rb^0} \ \p_\theta {\Rb^0}, & D_z &=  A^{-3/2} \left( \left( \p_{z} {\Rb^0} \right)^2 - A \right).
\end{align}
For a base interface of the form $r=\Rb^0(t,\theta)$ and a perturbation ansatz~(\ref{eq:internaldrainage_eigenmode_ansatz}), the normal vector of the base interface can be further simplified as
\begin{equation} \label{eq:internaldrainage-normal-base2}
	{\textbf{n}}^0 = \begin{pmatrix} n^0_r   \\ n^0_\theta   \\  n^0_z \end{pmatrix} = A^{-1/2} \begin{pmatrix} 1   \\ -\frac{1}{\Rb^0} \p_{\theta} {\Rb^0}  \\  0 \end{pmatrix},
\end{equation}
and simplification of the linearised perturbation of the normal vector~(\ref{eq:internaldrainage-normal-linearized})-(\ref{eq:internaldrainage-normal-lin-coef}) implies
\begin{equation} \label{eq:internaldrainage-normal-linearized2}
	{\bf \tilde{n}} = \begin{pmatrix} \tilde{n}_r   \\ \tilde{n}_\theta   \\  \tilde{n}_z \end{pmatrix} =
	\begin{pmatrix} B_{r} \tilde{\eta} + C_{r} \p_\theta \tilde{\eta} \\ B_{\theta} \tilde{\eta} +  C_{\theta} \p_\theta \tilde{\eta}  \\   \mathrm{i}k D_{z} \tilde{\eta}  \end{pmatrix},
\end{equation}
where
\begin{align} \label{eq:internaldrainage-normal-lin-coef2}
	\nonumber A &= 1 + \left( \frac{1}{\Rb^0} \p_{\theta} {\Rb^0} \right)^2 , &B_r &=  \frac{A^{-3/2}}{\left({\Rb^0}\right)^3} \left( \p_\theta {\Rb^0} \right)^2,  \\
	\nonumber B_\theta &= A^{-3/2}\left( -\frac{1}{\left({\Rb^0}\right)^4} \left( \p_\theta {\Rb^0} \right)^3 + \frac{A}{\left({\Rb^0}\right)^2} \p_\theta {\Rb^0} \right), &C_r &=  - \frac{A^{-3/2}}{\left({\Rb^0}\right)^2} \p_\theta {\Rb^0}, \\
	C_\theta &= -\frac{A^{-3/2}}{{\Rb^0}} , & D_z &=  -A^{-1/2}.
\end{align}

\subsection{Curvature}\label{app:internaldrainage_interfacecharacterization_curvature}
We recall that the local curvature of a linearly perturbed interface can be decomposed as ${\bf \kappa} = {\bf \kappa}^0 + \epsilon {\bf \kappa}^1$ which reads
\begin{equation} \label{eq:internaldrainage-curvature-base}
	\kappa^0 = \frac{1}{\Rb^0} \left(  n^0_r + \p_\theta  n^0_\theta  \right)+ \p_z  n^0_z,
\end{equation}	
\begin{align} \label{eq:internaldrainage-curvature-linearized}
	\nonumber \kappa^1 & = - \frac{\eta^1}{\left({\Rb^0}\right)^2} \left( n^0_r + \p_\theta n^0_\theta \right) + \frac{1}{{\Rb^0}} \left( n^1_r + \p_\theta n^1_\theta \right) + \p_z n^1_z \\
	& = E \eta^1 + F \p_\theta  \eta^1 + G \p_z  \eta^1 + D_z \p^2_{zz} \eta^1 + \frac{C_\theta}{{\Rb^0}} \p^2_{\theta \theta} \eta^1,
\end{align}
where
\begin{align} \label{eq:internaldrainage-curvature-lin-coef}
	\nonumber E &= \frac{1}{\Rb^0} \left( B_r + \p_\theta B_\theta  \right) - \frac{1}{\left({\Rb^0}\right)^2} \left( n^0_r + \p_\theta n^0_\theta \right) + \p_{z} B_z , \\ 
	\nonumber F &= \frac{1}{\Rb^0} \left( C_r + \p_\theta C_\theta + B_\theta + D_\theta \p_{z}  \right) + \p_{z} C_z  + C_z \p_{z} , \\ 
	G &=  B_z + \p_{z} D_z.
\end{align}
Note that the subscript $r$ in~(\ref{eq:internaldrainage-normal-base})-(\ref{eq:internaldrainage-curvature-lin-coef}) does not imply the real part. For a base interface of the form $r=\Rb^0(t,\theta)$ and a perturbation ansatz~(\ref{eq:internaldrainage_eigenmode_ansatz}), the curvature of the base interface can be further simplified as
\begin{equation} \label{eq:internaldrainage-curvature-base2}
	\kappa^0 = \frac{1}{\Rb^0} \left(  n^0_r + \p_\theta  n^0_\theta  \right),
\end{equation}
and simplification of the linearised perturbation of the curvature~(\ref{eq:internaldrainage-curvature-linearized})-(\ref{eq:internaldrainage-curvature-lin-coef}) implies
\begin{align} \label{eq:internaldrainage-curvature-linearized2}
	\nonumber \tilde{\kappa} & = - \frac{\tilde{\eta}}{\left({\Rb^0}\right)^2} \left( n^0_r + \p_\theta n^0_\theta \right) + \frac{1}{{\Rb^0}} \left( \tilde{n}_r + \p_\theta \tilde{n}_\theta \right) + \mathrm{i}k \tilde{n}_z \\
	& = E \tilde{\eta} + F \p_\theta  \tilde{\eta} -k^2 D_z \tilde{\eta} + \frac{C_\theta}{{\Rb^0}} \p^2_{\theta \theta} \tilde{\eta},
\end{align}
where
\begin{align} \label{eq:internaldrainage-curvature-lin-coef2}
	\nonumber E &= \frac{1}{\Rb^0} \left( B_r + \p_\theta B_\theta  \right) - \frac{1}{\left({\Rb^0}\right)^2} \left( n^0_r + \p_\theta n^0_\theta \right) , \\ 
	\nonumber F &= \frac{1}{\Rb^0} \left( C_r + \p_\theta C_\theta + B_\theta \right).
\end{align}

 \section{Derivation of a simplified expression for the energy density of the transient response.}\label{app:tgmatrix} \FloatBarrier

The interfacial energy density per wavelength is proportional to

\begin{equation}
\begin{split}
e(T) &= \frac{k}{2 \pi} \int_{0}^{2\pi} \int_{0}^{2\pi/k} |\bn(T)e^{\mathrm{i}kz} + c.c |^2 \mathrm{d}z \Rb^{0}(T)\mathrm{d}\theta \\
&= 2 \int_{0}^{2\pi}|\bn(T)|^2 \Rb^{0}(T)\mathrm{d}\theta \\
&= 2   \sum_{m=-N}^{N} \sum_{n=-N}^{N} \al_m^{\star}\al_n \int_{0}^{2\pi}\bn_m(T)^{\star}\bn_n(T) \Rb^{0}(T)\mathrm{d}\theta \\
&= 2 \bm{\al}^H\mathsfbi{A}(T)\bm{\al}
\end{split}
\end{equation}
where the strictly positive definite and Hermitian matrix $\mathsfbi{A}(T)$ is such that 
\begin{equation}
[\mathsfbi{A}(T)]_{mn} = \int_{0}^{2\pi}\bn_m(T)^{\star}\bn_n(T) \Rb^{0}(T)\mathrm{d}\theta, 
\end{equation}
for $-N\leq m\leq N$ and $-N\leq n\leq N$, and where we defined
\begin{equation}
\begin{split}
\bm{\al} = [\al_{N}^{\star},\al_{N-1}^{\star},...,\al_{1}^{\star},\al_0,\al_{1},...,\al_{N-1},\al_{N}]^T.
\end{split}
\end{equation}
The leading eigenvector of $\mathsfbi{A}(T)$ has no particular reason to have its $N$ first elements equal to the complex conjugate of its last $N$. Thus, it cannot correspond directly to the optimal set of $\bm{\al}$, that should satisfy this last constraint. For this reason we introduce the matrix 
\begin{equation}
\begin{split}
\mathsfbi{M}^{-1} = \begin{bmatrix}
\mathsfbi{I}& \mathsfbi{O} & \mathsfbi{P}\\ 
\mathsfbi{O}^T & \sqrt{2} & \mathsfbi{O}^T\\ 
-\mathrm{i}\mathsfbi{I} & \mathsfbi{O} & \mathrm{i}\mathsfbi{P}
\end{bmatrix}, \quad \text{with}  \quad \mathsfbi{I}_{ij} = \delta_{ij}, \quad  \mathsfbi{P}_{ij} = \delta_{N-i+1,j}, \quad \text{and}  \quad \mathsfbi{O}_i = 0.
\end{split}
\end{equation}
for $1\leq i,j\leq N$. Namely, $\mathsfbi{I}$ is the identity matrix of size $N\times N$, $\mathsfbi{P}$ is the identity matrix mirrored around its vertical axis (it contains ones on the diagonal from the bottom-left to the top-right and zeros everywhere else) and $\mathsfbi{O}$ is a vectors of zeros of size $N\times 1$. In this manner, we have directly $\bm{a} =  \mathsfbi{M}^{-1}\bm{\al}$, such that : 
\begin{equation}
\begin{split}
e(T) &= 2 \bm{\al}^H\mathsfbi{A}(T)\bm{\al} = 2 \bm{a}^T\mathsfbi{M}^H\mathsfbi{A}(T)\mathsfbi{M}\bm{a} =  2 \bm{a}^T\Re[\mathsfbi{M}^H\mathsfbi{A}(T)\mathsfbi{M}]\bm{a} 
\end{split}
\end{equation}

where we used that $\bm{a}$ and $e(T)$ are a real-valued. Therefore, defining $\mathsfbi{E}(T)=\Re[\mathsfbi{M}^H\mathsfbi{A}(T)\mathsfbi{M}]$ leads to the desired result.

 \section{Validation of the numerical model}\label{app:internaldrainage_validation} \FloatBarrier

\begin{figure}
	\centerline{ \includegraphics[width=1\textwidth]{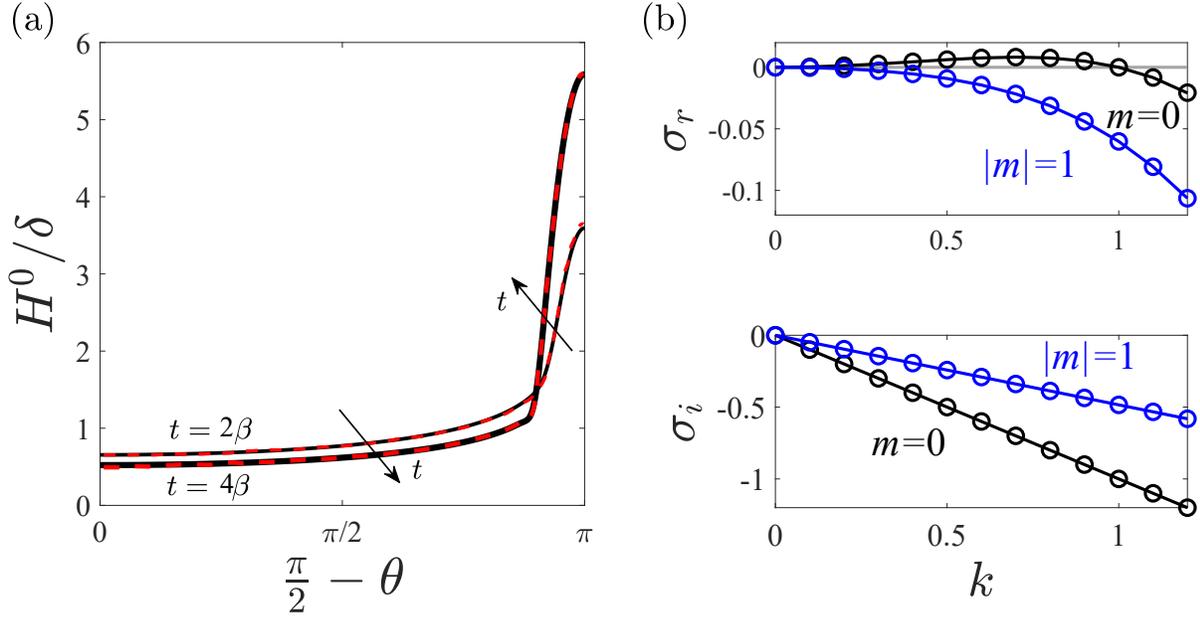} } 
	\caption{Numerical model validation; (a) Base flow: Two snapshots of the liquid film thickness, $H^0 = \beta - \Rb^0$, in the right half-plane, at time $t=\{2\beta, 4\beta \}$. Solid lines present the solution from the present study, and the dashed lines present the solution obtained by~\citet{Balestra2016}; $Oh =10.05, Bo=1960.2, \delta/ \beta=0.01$; (b) Stability analysis: dispersion curves of the two least stable modes associated with the viscous film coating inside a vertical tube, namely $|m|=\{0,1\}$. The continuous lines present the analytical solution obtained from the Stokes equations, and the circle represents the results from the present model; $Oh \rightarrow \infty, Bo=1, \delta=0.1$.}
	\label{fig:internaldrainage-model-validation}
\end{figure}
The developed numerical scheme is validated hereafter. Several measures are taken to ensure the correspondence of the model, based on the asymptotic limits where an analytical solution may exist.
\subsection{Base flow model}\label{app:internaldrainage_validation_BF}
The present base flow model is validated with \citet{Balestra2016} who studied a similar base flow in the limit of thin film, $\delta \ll 1$, and small surface tension, $Bo \gg 1$, by employing lubrication equations~\citep{Oron1997}. Figure~\ref{fig:internaldrainage-model-validation}(a) shows that the present model gives a solution of the base flow in full agreement with the solution of~\citet{Balestra2016}. 
\subsection{Linear stability analysis model}\label{app:internaldrainage_validation_LSA}
The present linear stability model is validated with the analytical solution that~\citet{Camassa2014} presented for the gravity-driven flow of a viscous film that coats the interior side of a vertical tube (where gravity points in $z$ direction in figure~\ref{fig:internaldrainage_schematic}). The corresponding base flow is parallel and can be expressed in cylindrical coordinates as 
\begin{equation} \label{eq:drainage-Nusselt-flow}
	u^0_z= \frac{\delta^{-2}}{2}  \left( \frac{\beta^2 - r^2}{2} + \ln{\frac{r}{\beta}} \right), \quad
	p^0 = -1, \quad
	\Rb^0 = 1,
\end{equation}
For the linear stability analysis,~\citet{Camassa2014} employed an approximation of the Stokes equations for long jets, referred to as the {\it long-wavelength approximation}~\citep{Reynolds1886} and compared the results with the analytical solution, in terms of Bessel functions, obtained by solving the full Stokes equations~\citep{Goren1962}. Thanks to the axisymmetry of the base flow,~\citet{Camassa2014} considered a perturbation as in (\ref{eq:internaldrainage_eigenmode_ansatz}) with the Fourier ansatz exponent of $\exp[\sigma t + \mathrm{i}kz + \mathrm{i}m\theta]$ where $m$ denotes the azimuthal wavenumber. Both of the aforementioned equations match in the limit of thin liquid film, $\beta \rightarrow 1$, where the long-wavelength approximation gives a dispersion relation of $\sigma = S \left(k^2-k^4\right) - \mathrm{i} k$ with $S=S(Bo,\delta)$ being constant for the axisymmetric perturbation $m=0$. All of the helical perturbations, $|m|>0$, are known to be linearly stable for such a flow~\citep{Rayleigh1878}. Figure~\ref{fig:internaldrainage-model-validation}(b) presents the full agreement between the present linear stability model and the analytical solution for a relatively thin film thickness $\delta=0.1$. It should be noted that despite the axisymmetric nature of the validated case, this presented validation holds also for an arbitrary interface. For this aim, the geometrical symmetry in the numerical reference frame is broken by setting the origin of the coordinates system at an arbitrary location inside the bubble, $(x,y)=(0.2,0.7)$. 
  
\subsection{Grid independency}\label{app:internaldrainage_validation_mesh}
\begin{figure}
		\centerline{\includegraphics[width=1\textwidth]{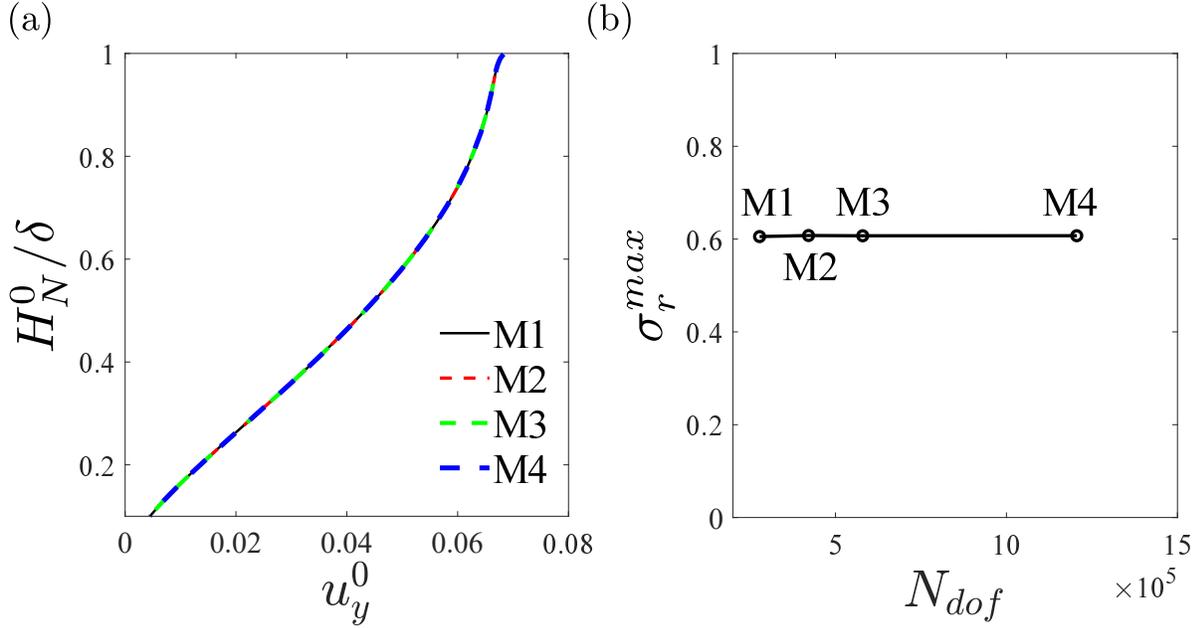}}
	\caption{Mesh convergence proof for $Oh \rightarrow \infty, Bo=0.05, \delta / \beta=0.3$: (a) base flow; $H_N^0 / \delta$ vs $u_y^0$; (b) linear stability analysis; $\sigma_r^{max}$ vs $N_{dof}$. All of the results presented in this manuscript are obtained for M3.}
	\label{fig:internaldrainage_mesh_dep}
\end{figure}

\indent A convergence study for the base flow evolution and the linear stability of the unstable eigenvalue is presented in figure~\ref{fig:internaldrainage_mesh_dep}, for $\{ Oh \rightarrow \infty, Bo=0.05, \delta / \beta=0.3 \}$. Mesh resolution is controlled by setting the number of divisions on the solid wall and interface boundaries. Mesh convergence is already attained for the presented grids. All of the presented results in the manuscript are obtained employing M3.



\begin{thebibliography}{99}
		
		\expandafter\ifx\csname natexlab\endcsname\relax
		\def\natexlab#1{#1}\fi
		\expandafter\ifx\csname selectlanguage\endcsname\relax
		\def\selectlanguage#1{\relax}\fi

		\bibitem[Augello {\it {et al.}} (2018)]{Augello2018}
		{\sc Augello, L., Fani, A. \& Gallaire, F.} 2018 {The influence of the entry region on the instability of a coflowing injector device}.  {\it J. Phys.: Condens. Matter} {\bf 30}, 284003, {\url{https://doi.org/10.1088/1361-648X/aac790}}.

		
		\bibitem[Balestra {\it {et al.}} (2016)]{Balestra2016}
		{\sc Balestra, G., Brun, P.-T. \& Gallaire, F.} 2016 {Rayleigh-Taylor instability under curved substrates: An optimal transient growth analysis}.  {\it Phys. Rev. Fluids} {\bf 1}(8), 083902, {\url{https://doi.org/10.1103/PhysRevFluids.1.083902}}.
		
		\bibitem[Balestra {\it {et al.}} (2018)]{Balestra2018}
		{\sc Balestra, G., Kofman, N., Brun, P.-T., Scheid, B. \& Gallaire, F.} 2018 {Three-dimensional Rayleigh--Taylor instability under a unidirectional curved substrate}. {\it J. Fluid Mech.} {\bf 837}, 19--47, {\url{https://doi.org/10.1017/jfm.2017.817}}.

        \bibitem[Balestra {\it {et al.}} (2018)]{Balestra20182}
		{\sc Balestra, G., Nguyen, D. M.-P. \& Gallaire, F.} 2018 {Rayleigh-Taylor instability under a spherical substrate}. {\it Phys. Rev. Fluids} {\bf 3} (8), 084005, {\url{https://link.aps.org/doi/10.1103/PhysRevFluids.3.084005}}.


        \bibitem[Balestra {\it {et al.}} (2018)]{Balestra20183}
		{\sc Balestra, G., Zhu, L. \& Gallaire, F.} 2018 {Viscous Taylor droplets in axisymmetric and planar tubes: From Bretherton’s theory to empirical models}. {\it Microfluid. Nanofluid.} {\bf 22}, 67, {\url{https://doi.org/10.1007/s10404-018-2084-y}}.

        \bibitem[Benilov (2006)]{Benilov2006}
		{\sc Benilov, E. S.} 2006 {Does surface tension stabilize a liquid film inside a rotating horizontal cylinder? Part 2: Multi-dimensional disturbances}. {\it Stud. Appl. Maths} {\bf 116}, 1--20, {\url{ https://doi.org/10.1111/j.1467-9590.2005.00331.x}}.
	
		\bibitem[Benilov {\it {et al.}} (2005)]{Benilovetal2005}
		{\sc Benilov, E. S., Kopteva, N. \& O’Brien, S. B. G.} 2005 {Does surface tension stabilize a liquid film inside a rotating horizontal cylinder?}. {\it Q. J. Mech. Appl. Maths} {\bf 58}, 158–-200, {\url{ https://doi.org/10.1093/qjmamj/hbi004}}.
		

		
		\bibitem[Bian {\it {et al.}} (2010)]{Bian2010}
		{\sc Bian, S., Tai, C-F., Halpern, D., Zheng, Y. \& Grotberg, J.B.} 2016 {Experimental study of flow fields in an airway closure model}. {\it J. Fluid Mech.} {\bf 647}, 391--402, {\url{https://doi.org/10.1017/S0022112010000091}}.


		\bibitem[Camassa {\it {et al.}} (2014)]{Camassa2014}
		{\sc Camassa, R., Ogrosky, H. R. \& Olander J.} 2014 {Viscous film flow coating the interior of a vertical tube. Part 1. Gravity-driven flow}. {\it J. Fluid Mech.} {\bf 745}, 682--715, {\url{https://doi.org/10.1017/jfm.2014.90}}.
	
		\bibitem[Camassa {\it {et al.}} (2016)]{Camassa2016}
		{\sc Camassa, R., Marzuola, J., Ogrosky, H. R. \& Vaughn, N.} 2016 {Traveling waves for a model of gravity-driven film flows in cylindrical domains}. {\it Phys. D} {\bf 333}, 254--265, \url{https://doi.org/10.1016/j.physd.2015.12.003}.
		
		\bibitem[Camassa {\it {et al.}} (2017)]{Camassa2017}
		{\sc Camassa, R., Ogrosky, H. R. \& Olander J.} 2017 {Viscous film-flow coating the interior of a vertical tube. Part 2. Air-driven flow}. {\it J. Fluid Mech.} {\bf 825}, 1056--1090, {\url{https://doi.org/10.1017/jfm.2017.409}}.

        \bibitem[Del Guercio {\it {et al.}} (2014)]{DelGuercio2014}
		{\sc Del Guercio, G. and Cossu, C. and Pujals, G.} 2014 {Optimal streaks in the circular cylinder wake and suppression of the global instability}. {\it Journal of Fluid Mechanics}, {\bf 752}, 572-588.

		\bibitem[Ding {\it {et al.}} (2018)]{Ding2018}
		{\sc Ding, Z., Liu, R., Wong, T. N. \&  Yang, C.} 2018 {Absolute instability induced by Marangoni effect in thin liquid film flows on vertical cylindrical surfaces}. {\it Chem. Eng. Sc.}, {\bf 177}, 261--269, {\url{https://doi.org/10.1016/j.ces.2017.11.039}}.

 
        \bibitem[Dobson \& Chato (1998)]{Dobson1998}
		{\sc Dobson, M. K. \& Chato, J. C.} 1998 {Condensation in Smooth Horizontal Tubes}. {\it ASME. J. Heat Transfer}, {\bf 120}(1), 193-–213, {\url{https://doi.org/10.1115/1.2830043}}.
		

		\bibitem[Duclaux {\it {et al.}} (2006)]{Duclaux2006}
		{\sc Duclaux, V., Clanet, C. \& Qu{\'e}r{\'e}, D.} 2006 {The effects of gravity on the capillary instability in tubes}. {\it J. Fluid Mech.}, {\bf 556}, 217--226, {\url{https://doi.org/10.1017/S0022112006009505}}.
		
	\bibitem[Eggers \& Villermaux (2008)] {Eggers2008}
	   {\sc Eggers, J. \& Villermaux, E.} 2008 {Physics of liquid jets.} {\it Rep. Prog. Phys.}, {\bf 71}, 036601, {\url{http://stacks.iop.org/RoPP/71/036601}}.

     \bibitem[Fermigier {\it {et al.}} (1987)]{Fermigier1992}
    	{\sc  Fermigier, M., Limat, L., Westfreid, J. E., Boudinet, P. \& Quilliet, C} 1992 {Two-dimensional patterns in Rayleigh–Taylor instability of a thin layer}. {\it J. Fluid Mech.}, {\bf 236}, 349–-383, {\url{https://doi.org/10.1017/S0022112092001447}}.


		\bibitem[Frenkel {\it {et al.}} (1987)]{Frenkel1987}
		{\sc  Frenkel, A. L., Babchin, A. J., Levich, B. G., Shlang, T. \& Sivashinsky, G. I.} 1987 {Annular flows can keep unstable films from breakup: Nonlinear saturation of capillary instability}. {\it J. Colloid Interface Sci.}, {\bf 115} 1, 225--233, {\url{https://doi.org/10.1016/0021-9797(87)90027-0}}.

		\bibitem[Gallaire \& Brun (2017)]{Gallaire2017}
		{\sc Gallaire, F. \& Brun, P.-T. } 2017 {Fluid dynamic instabilities: theory and application to pattern forming in complex media}. {\it Phil. Trans. R. Soc. A}, {\bf 375} (2093),  20160155, {\url{https://doi.org/10.1098/rsta.2016.0155}}.
  
		\bibitem[Goldsmith \& Mason (1963)]{Goldsmith1963}
		{\sc Goldsmith, H. L. \& Mason, S. G.} 1963 {The flow of suspensions through tubes. II. Single large bubbles}. {\it J. Colloid Interface Sci.}, {\bf 18}(3), 237--261, {\url{https://doi.org/10.1016/0095-8522(63)90015-1}}.		
		
		\bibitem[Goren (1962)]{Goren1962}
		{\sc Goren, S.} 1962 {The instability of an annular thread of fluid}. {\it J. Fluid Mech.}, {\bf 12}(2), 309--319, {\url{https://doi.org/10.1017/S002211206200021X}}.
		
	    \bibitem[Halpern \& Grotberg (2003)]{Halpern2003}
		{\sc Halpern, D. \& Grotberg, J.} 2003 {Nonlinear saturation of the Rayleigh instability due to oscillatory flow in a liquid-lined tube}. {\it J. Fluid Mech.}, {\bf 492}, 251--270, {\url{https://doi.org/10.1017/S0022112003005573}}.
	
		\bibitem[Heil {\it {et al.}} (2008)]{Heil2008}
		{\sc Heil, M., Hazel, A. L. \& Smith, J. A.} 2008 {The mechanics of airway closure}. {\it Respir. Physiol. Neurobiol.}, {\bf 163}(1--3), 214--221, {\url{https://doi.org/10.1016/j.resp.2008.05.013}}.
		
		\bibitem[Hu \& Cubaud (2020)]{Hu2020}
		{\sc Hu, X. \& Cubaud, T.} 2020 {From droplets to waves: Periodic instability patterns in highly viscous microfluidic flows}. {\it J. Fluid Mech.} {\bf 887}, A27, {\url{https://doi.org/10.1017/jfm.2019.1009}}.


      \bibitem[Joseph {\it {et al.}} (1997)]{Joseph1997}
		{\sc Joseph, D. D., Bai, R., Chen, K. \& Renardy, Y. Y.} 1997 {Core–annular flows}. {\it Ann. Rev. Fluid Mech.} {\bf 29}, 65--90, {\url{https://doi.org/10.1017/jfm.2019.1009}}.

            \bibitem[Joseph \& Renardy (1993)]{Joseph1993}
		{\sc Joseph, D. \& Renardy, Y.} 1993 {Fundamentals of Two-Fluid Dynamics, Part II}. {\it Springer}.

  
        \bibitem[Levy {\it {et al.}} (2014)]{Levy2014}
		{\sc Levy, R., Hill, D. B, Forest, M. G. \& Grotberg, J. B.} 2014 {Pulmonary Fluid Flow Challenges for Experimental and Mathematical Modeling}. {\it Integr. Comp. Biol.}, {\bf 54}(6), 985-–1000, {\url{https://doi.org/10.1093/icb/icu107}}.

    	\bibitem[Liu \& Ding (2017)]{Liu2017}
		{\sc Liu, R. \& Ding, Z.} 2017 {Stability of viscous film flow coating the interior of a vertical tube with a porous wall}.  {\it Phys. Rev. E} {\bf 95}(5), 053101, {\url{https://link.aps.org/doi/10.1103/PhysRevE.95.053101}}.		
		
		\bibitem[Ogrosky (2021)]{Ogrosky2021v}
		{\sc Ogrosky, H. R.} 2021 {Impact of viscosity ratio on falling two-layer viscous film flow inside a tube}. {\it Phys. Rev. Fluids}, {\bf 6}, 104005, {\url{https://doi.org/10.1103/PhysRevFluids.6.104005}}.
		
        \bibitem[Ogrosky (2021)]{Ogrosky2021}
		{\sc Ogrosky, H. R.} 2021 {Linear stability and nonlinear dynamics in a long-wave model of film flows inside a tube in the presence of surfactant}. {\it J. Fluid Mech.}, {\bf 908}, A23, {\url{https://doi.org/10.1017/jfm.2020.878}}.
				
		\bibitem[O'Neill \& Mudawar (2020)]{O'Neill2020}
		{\sc O'Neill, L. \& Mudawar, I.} 2020 {Review of two-phase flow instabilities in macro- and micro-channel systems}. {\it Intl J. Heat Mass transfer}, {\bf 157}, 119738, {\url{https://doi.org/10.1016/j.ijheatmasstransfer.2020.119738}}.		
				
		\bibitem[Oron {\it {et al.}} (1997)]{Oron1997}
		{\sc Oron, A., Davis, S. H. \& Bankoff, S. G.} 1997 {Long-scale evolution of thin liquid films}. {\it Rev. Mod. Phys.} {\bf 69}(931), 931--980, {\url{https://doi.org/10.1103/RevModPhys.69.931}}.
	
		\bibitem[Plateau (1873)]{Plateau1873}
		{\sc Plateau, J. A. F.} 1873 {Statique exp{\'e}rimentale et th{\'e}orique des liquides soumis aux seules forces mol{\'e}culaires}. {\it Gauthier-Villars}.

  
		\bibitem[Revankar \& Pollock (2005)]{Revankar2005}
		{\sc Revankar, S. T. \& Pollock, D. } 2005 {Laminar film condensation in a vertical tube in the presence of noncondensable gas}. {\it Appl. Math. Model.}, {\bf 29}(4), 341--359, {\url{https://doi.org/10.1016/j.apm.2004.09.010}}.
	
		\bibitem[Reynolds (1886)]{Reynolds1886}
		{\sc Reynolds, O.} 1886 {IV. On the theory of lubrication and its application to Mr. Beauchamp tower’s experiments, including an experimental determination of the viscosity of olive oil}. {\it Phil Trans. R. Soc. Lond.}, {\bf 177}, 157-234.

  
		\bibitem[Rayleigh (1878)]{Rayleigh1878}
		{\sc Rayleigh, L.} 1878 {On the instability of jets}. {\it Proc. Lond. Math. Soc.}, {\bf s1-10}, 4-13, {\url{https://doi.org/10.1112/plms/s1-10.1.4}}.	

        \bibitem[Rayleigh (1882)]{Rayleigh1882}
		{\sc Rayleigh, L.} 1878 {Investigation of the character of the equilibrium of an incompressible heavy fluid of variable density}. {\it Proc. Lond. Math. Soc.}, {\bf s1-14}, 170--177, {\url{https://doi.org/10.1112/plms/s1-14.1.170}}.	


        \bibitem[Tan \& Homsy (1986)]{Tan1986}
		{\sc Tan, C. T. \& Homsy, G.} 1986 {Stability of miscible displacements in porous media: Rectilinear flow}. {\it Phys. Fluids}, {\bf 29}(11), 3549--3556.

		\bibitem[Taylor (1950)]{Taylor1950}
		{\sc Taylor, G. I.} 1950 {The instability of liquid surfaces when accelerated in a direction perpendicular to their planes. I}. {\it Proc. R. Soc. Lond.}, {\bf A201}, 192-–196, {\url{https://doi.org/10.1098/rspa.1950.0052}}.


		\bibitem[Teng {\it {et al.}} (1999)]{Teng1999}
		{\sc Teng, H., Cheng, P.  \& Zhao, T. S.} 1999 {Instability of condensate film and capillary blocking in small-diameter-thermosyphon condensers}. {\it Heat Mass Transfer}, {\bf 42}, 3071--3083, {\url{https://doi.org/10.1016/S0017-9310(98)90375-1}}.		
		
        \bibitem[Trinh {\it {et al.}} (2014)]{Trinh2014}
		{\sc Trinh, P. H., Kim, H., Hammoud, N., Howell, P. D., Chapman, S. J. \& Stone, H. A.} 2014 {Curvature suppresses the Rayleigh-Taylor instability}. {\it Phys. Fluids} {\bf 26}(5), 051704, {\url{https://doi.org/10.1063/1.4876476}}.
		
	
	\end{thebibliography}

\end{document}